\documentclass[acmtog,nonacm]{acmart}
\usepackage[utf8]{inputenc}

\acmSubmissionID{1042}

\setcopyright{acmcopyright}
\acmJournal{TOG}
\acmYear{2025}
\acmVolume{45}
\acmNumber{6}
\acmArticle{1}
\acmMonth{12} 

\usepackage[utf8]{inputenc}
\usepackage[ruled,vlined,linesnumbered]{algorithm2e}   
\usepackage{listings}                           
\usepackage{newtxmath}                          
\usepackage{mathdots}
\usepackage{mathtools}
\usepackage{subcaption}
\usepackage{adjustbox}
\usepackage{multirow}
\usepackage{xspace}

\usepackage {enumitem}
\usepackage{tikz}
\usetikzlibrary{arrows.meta}                                                    
\usetikzlibrary{calc}
\newlength{\unit}
\newcommand\scalemath[2]{\scalebox{#1}{\mbox{\ensuremath{\displaystyle #2}}}}
\usepackage{arydshln}                                                           
\usepackage{calc}

\newlength{\gap}
\newlength{\Zheight}
\usepackage{xr}
\newcommand{\FLIP}{\protect\reflectbox{F}LIP\xspace}

\newcommand{\Zsymbol}{%
    \tikz[baseline=(C), x=\Zheight, y=\Zheight]{ 
        \coordinate (A) at (0, 1); 
        \coordinate (B) at (1, 1); 
        \coordinate (C) at (0, 0); 
        \coordinate (D) at (1, 0); 

        \node (base) at (0.5, 0.5) {}; 

        \foreach \point in {A, D} {
            \fill (\point) circle (0.25\Zheight); 
        }

        \foreach \point in {B, C} {
            \fill[gray] (\point) circle (0.25\Zheight); 
        }

        \draw[line width=0.1\Zheight] (A) -- (B);
        \draw[line width=0.1\Zheight] (C) -- (D);

        \draw[<->, line width=0.1\Zheight, >={Latex[scale=0.6]}] (B) -- (C);
    }%
}

\newcommand{\Ssymbol}{%
    \tikz[baseline=(C), x=\Zheight, y=\Zheight]{ 
        \coordinate (A) at (0, 1); 
        \coordinate (B) at (1, 1); 
        \coordinate (C) at (0, 0); 
        \coordinate (D) at (1, 0); 

        \node (base) at (0.5, 0.5) {}; 

        \foreach \point in {A, D} {
            \fill[gray] (\point) circle (0.25\Zheight); 
        }

        \foreach \point in {B, C} {
            \fill (\point) circle (0.25\Zheight); 
        }

        \draw[line width=0.1\Zheight] (A) -- (B);
        \draw[line width=0.1\Zheight] (C) -- (D);

        \draw[<->, line width=0.1\Zheight, >={Latex[scale=0.6]}] (A) -- (D);
    }%
}
\newcommand{\one}{black}
\newcommand{\zero}{black!10}
\newcommand{\Brick}[4]{%
    \tikz[baseline=(base), x=\Zheight, y=\Zheight]{ 
        \coordinate (base) at (0, 0); 
        \fill[#1] (0, 1) circle (0.4\Zheight); 
        \fill[#2] (1, 1) circle (0.4\Zheight);
        \fill[#3] (0, 0) circle (0.4\Zheight); 
        \fill[#4] (1, 0) circle (0.4\Zheight); 
    }%
}

\newlength{\matrixEntryWidth}
\newlength{\matrixBaselineShift}

\ExplSyntaxOn
\NewDocumentCommand{\binaryMatrix}{o m m} {
	\pgfmathsetmacro{\baselineAdjustment}{\IfNoValueTF{#1}{0}{#1}}
    \pgfmathsetmacro{\matrixSize}{#2}                                           
    \pgfmathsetmacro{\dotSize}{0.4}
    \pgfmathsetmacro{\centerY}{-0.5 * (\matrixSize - 1) + \baselineAdjustment}
    \tl_set:Nx \l_tmpa_tl { #3 }                                                
    \tl_remove_all:Nn \l_tmpa_tl { ~ }                                          
    \begin{tikzpicture}[
    	baseline=\centerY \matrixEntryWidth,
    	x = \matrixEntryWidth,
    	y = \matrixEntryWidth
    ]
        \pgfmathsetmacro{\x}{0}
        \pgfmathsetmacro{\y}{0}
        \tl_map_inline:Nn \l_tmpa_tl {
            \pgfmathsetmacro{\value}{int(##1 == "1")}                           
            \ifnum\value=1
                \fill[\one] (\x,\y) circle (\dotSize);                          
            \else
                \fill[\zero] (\x,\y) circle (\dotSize);                         
            \fi
            \pgfmathsetmacro{\x}{int(\x + 1)} 
            \ifnum\x = \matrixSize 
                \pgfmathsetmacro{\x}{0}
                \pgfmathsetmacro{\y}{\y - 1}
            \fi
        }
    \end{tikzpicture}
}

\ExplSyntaxOff

\definecolor{clrEmphasized}{rgb}{0.0, 0.18, 0.39}   

\author{Abdalla G. M. Ahmed}
\orcid{0000-0002-2348-6897}
\affiliation{%
  \institution{Shenzhen University}
  \country{China}
}
\email{abdalla_gafar@hotmail.com}

\author{Matt Pharr}
\orcid{0000-0002-0566-8291}
\affiliation{%
  \institution{NVIDIA}%
  \country{USA}}
\email{matt@pharr.org}

\author{Victor Ostromoukhov}
\affiliation{%
  \institution{Univ Lyon 1, CNRS, INSA Lyon}%
  \country{France}%
}
\email{victor.ostromoukhov@liris.cnrs.fr}

\author{Hui Huang}
\authornote{Corresponding Author.}
\affiliation{%
  \institution{Shenzhen University}
  \country{China}
}
\email{}

\title{SZ Sequences: Binary-Based $(0, 2^q)$-Sequences}

\begin{abstract}
Low-discrepancy sequences have seen widespread adoption in computer graphics thanks to the superior rates of convergence that they provide.
Because rendering integrals often are comprised of products of lower-dimensional integrals, recent work has focused on developing sequences that are also well-distributed in lower-dimensional projections. To this end, we introduce a novel construction of binary-based $(0, 4)$-sequences; that is, progressive fully multi-stratified sequences of 4D points, and extend the idea to higher power-of-two dimensions. We further show that not only it is possible to nest lower-dimensional sequences in higher-dimensional ones---for example, embedding a $(0, 2)$-sequence within our $(0, 4)$-sequence---but that we can ensemble two $(0, 2)$-sequences into a $(0, 4)$-sequence, four $(0, 4)$-sequences into a $(0, 16)$-sequence, and so on. Such sequences can provide excellent rates of convergence when integrals include lower-dimensional integration problems in 2, 4, 16,$\ldots$ dimensions. Our construction is based on using 2$\times$2 block matrices as symbols to construct larger matrices that potentially generate a sequence with the target $(0, s)$-sequence in base $s$ property. We describe how to search for suitable alphabets and identify two distinct, cross-related alphabets of block symbols, which we call $s$ and $z$, hence \emph{SZ} for the resulting family of sequences.
Given the alphabets, we construct candidate generator matrices and search for valid sets of matrices. We then infer a simple recurrence formula to construct full-resolution (64-bit) matrices.
Because our generator matrices are binary, they allow highly-efficient implementation using bitwise operations and can be used as a drop-in replacement for Sobol matrices in existing applications. 
We compare SZ sequences to state-of-the-art low discrepancy sequences, and demonstrate mean relative squared error improvements up to $1.93\times$ in common rendering applications.

\end{abstract}

\begin{teaserfigure}
  {\centering
     \scriptsize%
     \includegraphics[width=\textwidth]{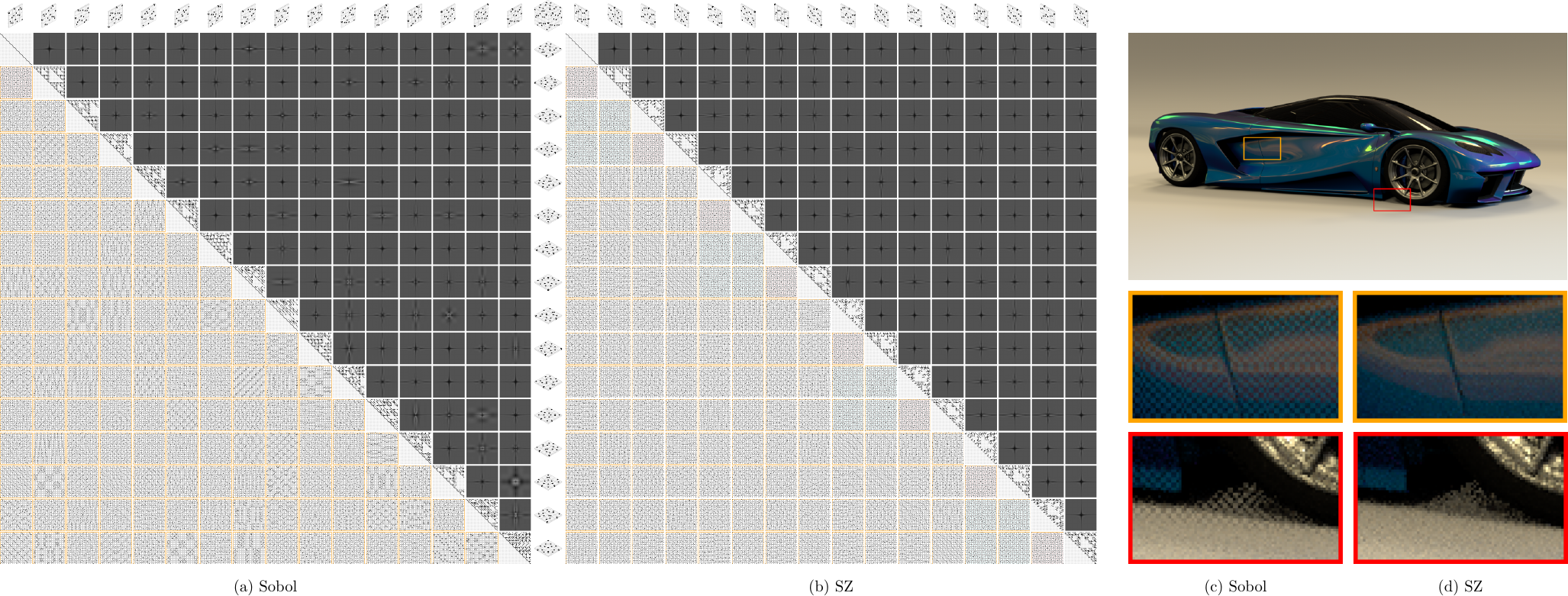}
     }%
     \caption{\label{fig:teaser}%
       (a) Visualization of all pairwise projections of the first 16 dimensions of the Sobol sequence.
       The binary Sobol generator matrices are shown along the diagonal.
       Underneath them, each square shows the first 256 points of the 2D projection using the pair of matrices in its row and column.
       Above the matrices are power spectra of the projection using the pair of matrices in the corresponding row and column, averaged over 1 million Owen-scrambled realizations.
       (b) The corresponding visualization for our SZ points.
       SZ points are well-distributed over all 2D projections; here we see that all sets of 256 points satisfy $16\times 16$ stratification, unlike Sobol points.
       The power spectra of their projections are much more regular and smoother at higher frequencies.
       Further, unlike Sobol, where only the first two dimensions are a $(0,2)$-sequence (red background),
       with SZ points each successive pair of matrices yields a $(0,2)$-sequence.
       Further, each consecutive set of 4 matrices generates 4D points that are $(0,4)$-sequences (green background), which means that they not only have all combinations of $4\times64$ stratifications in 2D, but also all combinations of $4\times 4\times 16$ stratifications for all 4 triplets in 3D, as illustrated in the exploded cube for the first three dimensions, and then there is a full $4\times 4\times 4\times 4$ stratification in 4D.
       Finally, all these 16 SZ matrices together give a $(0,16)$-sequence.
       (c, d) These properties lead to superior results in common rendering applications: the insets on the left are rendered using Sobol points
       and exhibit unconverged Sobol's characteristic checkerboard pattern while the insets on the right, using SZ points,
       do not suffer from this artifact, and have lower numeric error as well.
       For this scene, overall mean relative squared error is $1.93\times$ lower at 64 samples per pixel with SZ points.     
       }
\end{teaserfigure}

\begin{document}

%
%
%
\maketitle

\section{Introduction}
\label{sec:introduction}

Sampling is a fundamental process in computer graphics (CG), underlying multiple areas including halftoning and stippling, geometry processing, machine learning, and most notably rendering, where pixels are computed by Monte-Carlo integration of complex light-transport paths.
Even though these paths are inherently high-dimensional, they are mostly built from from 2D surface interactions, in addition to some 1D constituents; from here emerged the idea of similarly building the high-dimensional samples by pairing 2D and 1D samples over these constituents \cite{Cook1984Distributed,Glassner1995Principles}.
The research focus therefore went to optimizing 2D distributions, and it was long believed that optimizing over the high-dimensional space is ``at best a secondary concern'' \cite{Pharr04PBRT}.
Even when the inherently high-dimensional low-discrepancy (LD) sequences were brought to CG, generally only the first two dimensions were used and similarly ``padded'' to build high-dimensional samples \cite{Kollig02Efficient}.

While LD sequences could be used to sample the image plane directly~\cite{Mitchell1992ray}, doing so with parallel rendering based on screen-space decomposition was challenging since consecutive image samples end up at far-away pixels.
A milestone came with Gr{\"u}nschlo{\ss} et. al.~\shortcite{Gruenschlos12Enumerating} showing how to invert LD sequences, making it possible to determine which sample in the sequence corresponds to a given sample index in a pixel.
This inversion greatly facilitated \emph{global} LD sampling, where a single high-dimensional LD sequence is used to generate all samples for all pixels in all dimensions.
Despite noticeable artifacts before convergence, the undeniably superior performance of this sampling strategy clearly indicated that high-dimensional uniformity, in fact, does matter. It is worth noting that the employed Sobol sequences are already optimized for improved 2D projections by Joe and Kuo~\shortcite{Joe2008Constructing}, hence it is actually a combination of two- and high-dimensional uniformity that accounts for this success.
Soon after, \emph{pbrt} and Mitsuba, the common research rendering engines, both adopted global Sobol sampling for their defaults, and most of subsequent sampling research in CG was devoted to LD constructions, with a primary goal of imposing more control on these distributions, especially their lower-dimensional projections.

In this paper, we make a major advancement in this research direction of controlling LD distributions.
Rather than adapting one of the few known sequences, we present a novel construction of a complete family of LD sequences, built entirely from first principles.
We use binary matrices to construct (0, 4)-sequences in base 4, as defined in Section~\ref{sec:related work} below.
We then show how to extend this sequence to additional dimensions, achieving $(0,2)$-sequences in each pair of dimensions, $(0,4)$-sequences in each quad, $(0,16)$-sequences in each set of 16 dimensions, and so on.
Our sequences thus give superior integration performance in many cases where lower-dimensional projections are important.
Because our generator matrices are binary, they can be used as a drop-in replacement in systems that currently use Sobol samples;
no algorithmic changes are necessary, and performance is unaffected with their substitution.

\section{Technical Background and Related Work}
\label{sec:related work}

Our work belongs to the research area on uniform distribution of points \cite{Kuipers74Uniform}, aiming at obtaining point distributions that \emph{uniformly} cover a domain better than \emph{random} (white noise) distributions. The $s$-dimensional sampled domain is usually scaled to a half-open unit hypercube $(0, 1]^s$.
In this section, we furnish a basic technical background needed to understand our proposed method, and briefly review the most closely-related work.

\subsection{Discrepancy}

The term ``discrepancy'' refers to the error of using a point set to estimate the volume of region by counting the ratio of points falling inside.
Star discrepancy, commonly notated as $D^{*}$, refers to the maximum discrepancy measured over all axis-aligned hyper-rectangles in the concerned domain, and gives a reliable measure of the performance of the point set in numerical integration. A point set is considered a low-discrepancy (LD) set if it attains \mbox{$\mathcal{O}\left(\log^{s-1}(N)/N\right)$} discrepancy for $N$ points in $s$-dimensions.
A low-discrepancy sequence is an infinite sequence of points that provides an LD set for certain contiguous blocks of points of certain sizes, and attains \mbox{$\mathcal{O}\left(\log^{s}(N)/N\right)$} discrepancy for all contiguous blocks.

\subsection{Radix-Based Constructions}
\label{sec:radix-based}

A significant construction of a one-dimensional LD sequence was presented by van der Corput~\shortcite{vanderCorput1935}.
If 
\begin{equation}
    \ldots v_{3}v_{2}v_{1} = \sum_{i=1} v_i 2^{i-1}   \label{eq:binary expansion}
\end{equation}
denotes the binary encoding of the sample index $v$, then the $v$th sample is computed by mirroring the bits of $v$ at the fraction point:
\begin{equation}
    x_v = \sum_{i=1} v_i 2^{-i}\,,   \label{eq:vdC}
\end{equation}
This approach can be generalized to higher dimensions by generating the $k$th component/dimension of the sample using a distinct linear mapping
\begin{equation}
    X^{(k)}_v = C^{(k)} V\,,   \label{eq:digital construction}
\end{equation}
where $V$ is a vector representing the $v$th digit-reversed van-der-Corput-like sample in some base $b$, $X^{(k)}_v$ is a vector representing the same-base fractional digits of the final sample location along the $k$th axis, and $C^{(k)}$ is a matrix in Galois Field (GF) $b$, known as a \emph{generator matrix}, that linearly maps reversed digits of the sample index to digits of the computed sample, hence the name ``digital'' to collectively describe these methods. Next we briefly outline the three most established digital construction approaches to LD sequences.

The first known higher-dimensional extension to van der Corput sequences in due to Halton~\shortcite{Halton1960efficiency}, and simply uses a different prime base for encoding the index $v$ along each axis. Identity matrices may be used for $C^{(k)}$ in Eq.~\eqref{eq:digital construction}, but more complex  \emph{scrambling} matrices have been proposed to improve the pairwise 2D projections.

A very different approach was taken by Sobol~\shortcite{sobol1967}, who stuck to a binary-base encoding, and constructed a different generator matrix for each dimension, using a recurrence operation to compute each column of the matrix from the preceding columns, taking the coefficients from an irreducible polynomial. These polynomials, in a sense, replace the role of different bases in Halton's construction. We skip the details and only highlight the most relevant one here that the first two irreducible polynomials lead to the pair
\begin{equation}
    \setlength{\matrixBaselineShift}{-1.5pt}
    (I, P) = 
    \left(
    \binaryMatrix[\matrixBaselineShift]{32}{
        1............................... 
        .1.............................. 
        ..1............................. 
        ...1............................ 
        ....1........................... 
        .....1.......................... 
        ......1......................... 
        .......1........................ 
        ........1....................... 
        .........1...................... 
        ..........1..................... 
        ...........1.................... 
        ............1................... 
        .............1.................. 
        ..............1................. 
        ...............1................ 
        ................1............... 
        .................1.............. 
        ..................1............. 
        ...................1............ 
        ....................1........... 
        .....................1.......... 
        ......................1......... 
        .......................1........ 
        ........................1....... 
        .........................1...... 
        ..........................1..... 
        ...........................1.... 
        ............................1... 
        .............................1.. 
        ..............................1. 
        ...............................1 
    }\,,
    \binaryMatrix[\matrixBaselineShift]{32}{
        11111111111111111111111111111111
        .1.1.1.1.1.1.1.1.1.1.1.1.1.1.1.1
        ..11..11..11..11..11..11..11..11
        ...1...1...1...1...1...1...1...1
        ....1111....1111....1111....1111
        .....1.1.....1.1.....1.1.....1.1
        ......11......11......11......11
        .......1.......1.......1.......1
        ........11111111........11111111
        .........1.1.1.1.........1.1.1.1
        ..........11..11..........11..11
        ...........1...1...........1...1
        ............1111............1111
        .............1.1.............1.1
        ..............11..............11
        ...............1...............1
        ................1111111111111111
        .................1.1.1.1.1.1.1.1
        ..................11..11..11..11
        ...................1...1...1...1
        ....................1111....1111
        .....................1.1.....1.1
        ......................11......11
        .......................1.......1
        ........................11111111
        .........................1.1.1.1
        ..........................11..11
        ...........................1...1
        ............................1111
        .............................1.1
        ..............................11
        ...............................1
    }
    \right)  \label{eq:I-P}
\end{equation}
that constitutes the first two, invariant, and most well-known dimensions of Sobol's construction, where 
\begin{equation}
    P = \left(p_{i,j}: \binom{j}{i} \bmod 2 \right)
\end{equation}
happens to be the Pascal matrix in GF(2), where $i$/$j$ are row/column indices, indexed from 0.
Please note that we hereinafter adopt the convention of showing binary matrices as arrays of (0) gray and (1) black dots for better visibility, and also occasionally using dots instead of zeros for the same purpose.

These first two dimensions of Sobol inspired a third construction by Faure~\shortcite{Faure1982Discrepancy} that uses powers of a Pascal matrices
\begin{equation}
    C^{(k)} = \left(c^{(k)}_{i,j}:\binom{j}{i} k^{j-i} \bmod b \right)\,, \label{eq:Faure}
\end{equation}
in a prime base $b$, and generates a $b$-dimensional LD sequence analogous to the 2D Sobol sequence, in a sense explained in the following subsection.
Niederreiter~\shortcite{Niederreiter87Point} extended Faure construction to power-of-prime bases $b^q$ using symbolic matrices and arithmetic over the GF($b^q$).

Thus, we have three well-established constructions approaches to high-dimensional LD sequences:
\begin{enumerate}
    \item Halton, using different prime bases.
    \item Sobol and variants, using binary-polynomial-based matrices.
    \item Faure and Niederreiter extension, using Pascal matrices over fields.
\end{enumerate}
Among the three, two factors made Sobol the most adopted one in computer graphics: a) the binary base, leading to extremely fast computation, and b) that higher distribution quality is obtained for lower dimensions, as will be explained in the following subsection.
Halton comes next, sharing the second but not the first property of Sobol.
Because the prime bases are known at compile-time, the cost of integer divisions and modulus operations can be reduced~\cite{Warren2012Hacker}, bringing the digit reversal of Halton to within tolerable limits; this has led to making Halton sequences being used in common rendering platforms like \emph{pbrt}~\cite{Pharr16PBRT}.

In contrast, we are unaware of any mention of Faure sequences and their derivatives in CG, and we may attribute this primarily to the large base used for all dimensions, requiring a considerably large number of points to attain the advertised uniformity. The lookup-based arithmetic of Niederreiter's extension exacerbate this by adding considerable computational overhead.

While our work is developed from first principles, the resulting construction may be considered as an extension to Faure's, specifically Niederreiter's extension, that addresses both limitations: we ensemble higher-dimensional sequences from lower-dimensional ones, and use GF(2) matrices like Sobol's. Thus, our construction effectively introduces the third category of LD sequences to CG.

\subsection{$(t, m, s)$-Nets and $(t, s)$-Sequences}

Niederreiter~\shortcite{Niederreiter87Point} also established a complete theoretical framework for developing and studying a large class of LD constructions, including the radix-based ones listed above. At the heart of this theory is so-called $(t, m, s)$-Nets in base $b$, which refers to a \emph{multi-stratified} distribution of $b^m$ points in an $s$-dimensional hypercube such that, for all possible stratification (slicing) ways of the domain into $b^{m-t}$ similar rectangular \emph{strata} (cells), we find exactly $b^t$ points in each stratum.
An illustration of the essence of a $(0, 4, 4)$-net in base 4, for example, is shown in Fig.~\ref{fig:teaser}.
Then come $(t, s)$-sequences in base $b$, which are infinite sequences of $s$-dimensional points such that, for all integer $m$, the first and all subsequent blocks of $b^m$ points comprise a $(t, m, s)$-net in base $b$.
While these definitions are independent of the construction method, the model possibly existed thanks to the existence of algebraic recipes for construction.
Specifically, the definition of $(t, m, s)$-nets translates directly into a simple condition on digital:
if we build a hybrid $m\times m$ matrix
\begin{equation}
    \scalemath{1} {
    H_{m_1,m_2,\ldots,m_s}
    =
    \left(
    \begin{array}{lcl}
        c^{(1)}_{0,0}   & \ldots & c^{(1)}_{0,m-1}   \\
        \vdots          & \vdots & \vdots          \\
        c^{(1)}_{m_1-1,0} & \ldots & c^{(1)}_{m_1-1,m-1} \\
        \hline
        \vdots          & \vdots & \vdots          \\
        \hline
        c^{(s)}_{0,0}   & \ldots & c^{(s)}_{0,m-1}   \\
        \vdots          & \vdots & \vdots          \\
        c^{(s)}_{m_s-1,0} & \ldots & c^{(s)}_{m_s-1,m-1} \\
    \end{array}
    \right)\,,
    }
    \label{eq:hybrid matrix}
\end{equation}
by combining the first $m_k$ rows of the $k$th generator matrix in Eq.~\eqref{eq:digital construction}, with $m_k$s summing to $m$, then each such a hybrid matrix multiplied by $V$ computes the allocation of the points to strata in a specific stratification. The $(0, m, s)$-net condition, for example, then corresponds to requiring that each hybrid matrix is invertible, ensuring a one-to-one correspondence of points to strata. A $(0, s)$-sequence requires a set of generator matrices that maintains this hybrid matrix condition progressively over their top-left corners.

These definitions suggest that it is desirable to keep both the base $b$ and the parameter $t$ small, though different needs might call for different configurations.
To give some examples, the van der Corput construction is a (0, 1)-sequence in base 2. Faure sequences are $(0, b)$-sequences in their respective prime bases, while Niederreiter's extension to Faure creates $(0, b^q)$-sequences in power-of-prime bases $b^q$. The first two dimensions of Sobol sequence comprise a $(0, 2)$-sequence in base 2, but the $t$ value increases as we take more dimensions, which makes lower dimensions attain better quality, as we mentioned in Sec.~\ref{sec:radix-based}.
Halton sequences, in contrast, do not readily fit within the $(t, s)$-sequences model, unless we adapt it to accept different bases. They still exhibit a multistratified structure, though. Hence, the common ground of all the three families of constructions is that they attain optimal quality at powers of the base, or product of bases in case of Halton, and here we can see the reason of superiority of Sobol and Halton over Faure for graphics applications where lower dimensions are usually more important.

In this paper we develop a novel construction of $(0, 2^q)$-sequences using binary GF(2) matrices.
Technically speaking, we are breaking two traditional rules of nets and sequences: that a composite number cannot directly be used as a base, and that one can not obtain a $(0, s)$-sequence with a base $b < s$.
We achieve this result, however, by combining the two violations: instead of using a base-4 directly, for example, we use $2\times 2$ binary matrices as building blocks to emulate base-4 matrices.
Please note that $b = s$ is implied in the following if no base is mentioned.

\subsection{Scrambling and Shuffling of LD Constructions}
\label{sec:scrambling and shuffling}

Beyond the algebraic recipes for constructing LD sets and sequences, the design space is greatly enriched by \emph{scrambling} techniques that can derived new valid nets/sequences from given ones.
Tezuka~\shortcite{Tezuka1994Generalization} showed that multiplying arbitrary lower-triangular matrices from left by the generator matrices produces valid generators, since doing so preserves that ranks of all hybrid matrices.
Owen~\shortcite{Owen95Randomly}, in contrast, introduced a post-processing technique, ``Owen scrambling'', that preserves the net/sequence structure.
In contrast to scrambling that applies to sample locations, the term \emph{shuffling} refers to reordering (permuting) the sample indices \cite{Faure02Another}.
As $(0, s)$-sequences, all these are applicable to our sequences, and we will use them in our development.

\subsection{LD Sampling in Computer Graphics}

The interest in LD construction has grown in the graphics community over the past decade, mostly devoted to finding handles to impose control over these distributions as commonly needed in graphics applications.
Ahmed et al.~\shortcite{Ahmed16LDBN} presented an original 2D low-discrepancy construction that enables imposing a blue noise spectrum, then
Perreir et al.~\shortcite{Perrier18Sequences} imposed a blue noise spectrum onto selected pairs of dimensions of a Sobol sequence.
While both methods compromised the discrepancy of the underlying distribution, they presented a proof of concept that encouraged further research.

In a different path, Christensen et al.~\shortcite{Christensen18Progressive} tried searching for dyadic (i.e., base-2) $(0, 2)$-sequences sequences with good spatial and spectral properties, and Pharr~\shortcite{Pharr19Efficient} presented a more efficient search, but Ahmed and Wonka then presented a theorem~\shortcite[Theorem~4.2]{Ahmed2021Optimizing} implying that the search space is actually confined to Owen-scrambled 2D Sobol, and soon the first group of authors, lead by Helmer~\shortcite{Helmer2021Stochastice},  developed an extremely fast implementation of Owen scrambling. We take this as an excellent example of the fast-paced learning cycles of the community to analyze and manipulate the inherently difficult-to-understand LD construction. In this paper we present an almost complete cycle from experimentation to elements of a theory.

In more recent years, the research on LD in CG reached the state of enriching the quasi-Monte-Carlo (QMC) literature with significant findings and novel constructions, including an efficient algorithm that spans the whole space of dyadic $(0, m, 2)$-nets \cite{Ahmed2021Optimizing}, a cascade of dyadic $(0, m, 2)$-nets over successive pairs of dimensions \cite{Paulin2021Cascaded}, a family of self-similar dyadic $(0, 2)$-sequences~\cite{Ahmed23DigitalSequences}, a gradient-decent optimization of discrete Owen scrambling \cite{Doignies2025Differentiable}, and a base-3 analog to Sobol sequences~\cite{Ostromoukhov2024QuadOptimized}.
Our current work is partially inspired by this last work of Ostromoukhov et al., where we thought of using base 4 instead to gain the advantage of binary computation.

\section{Exploration}
\label{sec:exploration}

In this section we report our pilot search for binary matrices that generate $(0, 2^q)$-sequences, starting from experimental exploration and gradually developing an analytical framework distilled from empirical findings.
Starting with four dimensions, our initial goal is to construct a $(0, 4)$-sequence by searching for a working set of generator matrices that linearly map a reverse-ordered vector of digits of the sample index into vectors of digits representing coordinates along the four axes, as in Eq.~\eqref{eq:digital construction}.
It is already established and well known that plain matrices and straightforward arithmetic in a composite base like 4 would not serve the purpose.
An intuition of this deficiency may be obtained by noting that a factor of the base, like 2 in base 4, would not preserve all the ranks of a multiplied sequence number digit modulo base, which drastically limits the opportunity of finding a working set of matrices.
Rather than using GF(4) symbolic arithmetic like Niederreiter~\shortcite{Niederreiter87Point}, our idea is to use plain GF(2) matrices and vectors, interpreting $2\times 2$ blocks of the generator matrices and pairs of bits of the multiplied $V$ vector as base-4 digits.
Our first key insight is that the set of $2\times 2$ invertible binary matrix blocks is capable of resolving all the 6 permutations of the list of non-zero numbers in a base-4 digit.
While this does not prove that building matrices from such blocks would yield a $(0,4)$-sequence, it is at least encouraging to explore.

We start with an exhaustive search over a small range to validate the concept.
Without loss of generality, we may assume the identity matrix $I$ for the first dimension, since it can be factored out from the right of the four matrices to permute the sequence number.
Our plan, then, is to progressively search for three binary square matrices that expand at a pace of 2 rows and columns and fulfill the hybrid-matrices condition (Eq.~\eqref{eq:hybrid matrix}) in each step.
That is, in each step, and for each dimension, we expand each matrix
\begin{equation}
    A_{m + 2} \leftarrow
        \left(
            \begin{array}{c|c}
                 A_{m} & C \\\hline
                 R & X
            \end{array}
        \right)\,,      \label{eq:extension}
\end{equation}
by adding an $m\times 2$ column $C$, a $2\times m$ row $R$, and a $2\times 2$ corner block $X$.
We will drop the suffix hereinafter where not deemed ambiguous.
We may always assume zeros for $R$ by factoring a lower-triangular Tezuka-scrambling matrix:
\begin{equation}
    \left(
        \begin{array}{c|c}
             A & C \\\hline
             R & X
        \end{array}
    \right)
    =
    \left(
        \begin{array}{c|c}
             I & 0 \\\hline
             R A^{-1} & I
        \end{array}
    \right)
    \left(
        \begin{array}{c|c}
             A & C \\\hline
             0 & R A^{-1} C + X
        \end{array}
    \right)\,, \label{eq:R AInv C + X}      
\end{equation}
where $I$ is an appropriately sized identity matrix.
Then we use the factoring
\begin{equation}
    \left(
        \begin{array}{c|c}
             A & C \\\hline
             0 & X
        \end{array}
    \right)
    =
    \left(
        \begin{array}{c|c}
             I & 0 \\\hline
             0 & X
        \end{array}
    \right)\,.       
    \left(
        \begin{array}{c|c}
             A & C \\\hline
             0 & I
        \end{array}
    \right)\,,
\end{equation}
to reveal that $X$ must be invertible.
This reduces the set for corner extension matrices from 16 to the following six:
\begin{equation}
    \scalemath{0.8}{
        \left\{
            \left(\begin{array}{cc} 1 & . \\ . & 1  \end{array}  \right),
            \left(\begin{array}{cc} 1 & 1 \\ . & 1  \end{array}  \right),
            \left(\begin{array}{cc} 1 & . \\ 1 & 1  \end{array}  \right),
            \left(\begin{array}{cc} . & 1 \\ 1 & .  \end{array}  \right),
            \left(\begin{array}{cc} 1 & 1 \\ 1 & .  \end{array}  \right),
            \left(\begin{array}{cc} . & 1 \\ 1 & 1  \end{array}  \right)
        \right\}
    }\,. \label{eq:2x2 blocks}
\end{equation}
We may further exclude the third and fifth by factoring out a lower-triangular Tezuka-scrambling matrix, reducing the set to only four:
\begin{equation}
    \scalemath{0.8}{
        \left\{
            \left(\begin{array}{cc} 1 & . \\ . & 1  \end{array}  \right),
            \left(\begin{array}{cc} 1 & 1 \\ . & 1  \end{array}  \right),
            \left(\begin{array}{cc} . & 1 \\ 1 & .  \end{array}  \right),
            \left(\begin{array}{cc} . & 1 \\ 1 & 1  \end{array}  \right)
        \right\}
    }\,, \label{eq:2x2 blocks reduced}
\end{equation}
which is more favorable as a power of 2.

We built an efficient exhaustive search algorithm by placing a whole matrix compactly in a 64-bit word, allowing for fast bit manipulation and for building lookup tables of invertible matrices. This enabled searching up to $8\times 8$ matrices, i.e., four extension steps.
The search confirmed the existence of multiple working sets of matrices, which encouraged us to conduct two important experiments.
We first considered \emph{nesting} a (0, 2)-sequence, namely 2D Sobol, in our (0, 4) set.
Towards that end, we enforced the extension of the second-dimension matrix to reproduce the Pascal matrix, and search for a complementing pair of matrices.
This successfully worked all the way to 8$\times$8 matrices, with multiple choices in each extension.
Next, we considered \emph{ensembling} two (0, 2)-sequences into a (0, 4)-sequence.
To achieve this, we used the formula recently exposed by Ahmed et al.~\shortcite{Ahmed23DigitalSequences}, telling that for any pair $\{U_x, U_y\}$ of upper-triangular generator matrices to make a (0, 2)-sequence, they have to satisfy 
\begin{equation}
    U_y U_x^{-1} = U_x U_y^{-1} = P\,.
\end{equation}
Thus, we conducted the search as with nesting, and in each extension filtered the results that satisfy this condition.
Once again, the experiment worked successfully all the way to the $8\times 8$ matrices.

\subsection{Initial Findings}

For all combinations of valid initial 2$\times$2 blocks of the three matrices, we consistently obtained 768 distinct valid sets for the first extension to 4$\times$4, and 384 distinct valid extension on each branch for subsequent extension to 6$\times$6 and 8$\times$8, which suggests an analytical construction model.
The fixed number of alternatives in each extension step indicates that the degrees of freedom are possibly limited to the top and bottom blocks of the extension columns.
Following this hint, we note that the corner extension block $X$ in Eq.~\eqref{eq:extension} is not involved in any of the hybrid matrices, hence comprises, for each of the three matrices, a free choice among the four options in Eq.~\eqref{eq:2x2 blocks reduced}.
This accounts for a factor of $4^3 = 64$, and we verified that fixing the extension corner blocks $X$ to identity reduces the list to 12/6 sets in the first/subsequent extension steps.
This leaves us with the degrees of freedom in the top row, which we analyze next.

Let
\begin{equation}
    \left( X_0 \, X_1 \, \ldots \, X_{j-1} \, X_j \right)
\end{equation}
denote the first block-row in one matrix $X$ of the three upon the $j$th extension, where each $X_i$ is a 2$\times$2 block matrix.
By constructing a hybrid matrix comprising $j$ rows of the first identity matrix and this row of $X$, we find that
\begin{equation}
    \left|
        \begin{array}{ccc|c}
                & I      &         &  \\\hline
            X_0 & \ldots & X_{j-1} & X_j
        \end{array}
    \right| = 1 \implies \left|X_j\right| = 1\,,      \label{eq:rule 0}
\end{equation}
leading to the conclusion that, for each of the three matrices, all $2\times 2$ blocks of the first block row have to be invertible.

Now consider taking a hybrid of the first $j-1$ rows of $I$ with the first row of each of two other matrices $A$ and $B$,
then we get
\begin{equation}
    \left|
        \begin{array}{ccc|cc}
                 & I      &                   &     \\\hline
             A_0 & \ldots & A_{j-2} & A_{j-1} & A_j \\
             B_0 & \ldots & B_{j-2} & A_{j-1} & A_j 
        \end{array}
    \right| = 1 \implies
    \begin{vmatrix} A_{j-1} & A_j \\ B_{j-1} & B_j\end{vmatrix} = 1\,.
\end{equation}
Since all four elements are invertible by virtue of Eq.~\eqref{eq:rule 0}, we can use our earlier ``\mbox{$R A^{-1} C + X$}'' reduction of Eq.~\eqref{eq:R AInv C + X} to get
\begin{equation}
    \left| B_{j-1} A_{j-1}^{-1} A_j + B_j \right| = 1\,,
\end{equation}
or
\begin{equation}
    \left| A_{j-1}^{-1} A_j + B_{j-1} ^{-1} B_j \right| = 1\,. \label{eq:symbols-addition-implicit}
\end{equation}
Let us define a symbol
\begin{equation}
    x_j = X_{j-1}^{-1} X_j\,,   \label{eq:symbol definition}
\end{equation}
to designate the product of the inverse of the $j-1$st by the $j$th entry in the first row of block-matrix $X$.
Then Eq~\eqref{eq:symbols-addition-implicit}, rewritten as
\begin{equation}
    \left| a_j + b_j \right| = 1\,, \label{eq:symbols-addition-explicit}
\end{equation}
says that the sum of corresponding symbols in the first row of each pair $\{A, B\}$ of generator matrices must be invertible, which also implies that the symbols at corresponding slots must be distinct for the three matrices.
Thus, for each slot in the first block row, the inserted block-matrix symbols must be
\begin{enumerate}
    \item Invertible.
    \item Distinct for each matrix.
    \item Have invertible sums.
\end{enumerate}
The population of six invertible 2$\times$2 matrices comprises two disjoint sets,
\begin{equation}
    (S, Z) = \left(\left\{
        \underbrace{\Brick{\one }{\zero}{\zero}{\one }}_{s_1},
        \underbrace{\Brick{\zero}{\one }{\one }{\one }}_{s_2},
        \underbrace{\Brick{\one }{\one }{\one }{\zero}}_{s_3}
    \right\},
    \left\{
        \underbrace{\Brick{\zero}{\one }{\one }{\zero}}_{z_1},
        \underbrace{\Brick{\one }{\zero}{\one }{\one }}_{z_2},
        \underbrace{\Brick{\one }{\one }{\zero}{\one }}_{z_3}
    \right\}\right)\,,    \label{eq:S and Z symbols}
\end{equation}
that satisfy these requirements, each comprising two diagonally-mirrored triangular elements, and a counter diagonal element that switches between them via addition, hence we call them $S$ (think \Ssymbol) and $Z$ (think \Zsymbol) to make it easier to recall which is which.
Hence comes the name SZ for these sequences.
We will also stick hereinafter to this ``S/Z'' convention in discriminating left/right-hand elements and operations, respectively.

We conclude this subsection by noting that these three simple rules comprise all the requirements for satisfying the hybrid matrix condition for the first $2\times 2$ block matrices. Further, they analogously extend to other powers of 2. Thus, constructing binary matrices for generating a ``$(0, 2, 2^q)$-net in base $2^q$'' reduces to finding an alphabet of $q\times q$-block matrices satisfying these rules. This name in Niederreiter's notation just means a set of $2^{2q}$ points in $2^q$ dimensions that is stratified in all \mbox{$2^{q-1}(2^{q} - 1)$} pairs of 2D projections; cf. Jarosz et. al.~\shortcite{Jarosz19Orthogonal}.
Next, we show how to extend the matrices to arbitrary sizes for $(0, 2^q)$-sequences in base $2^q$.

\subsection{Refinement}

Instead of defining the symbols in Eq.~\eqref{eq:symbol definition} in terms of adjacent entries, it would be more constructive to populate the matrices directly with symbols.
Realizing that all first-row blocks must be invertible helps us to achieve this via this subtle factoring
\begin{align}
    &\scalemath{0.6}{
        \left(
    	    \begin{pmatrix}
    	        I &        &     \\
    	          & \ddots &  \\
    	          &        & I
    	    \end{pmatrix},
            \begin{pmatrix}
                I & \ldots & A_j    \\
                  & \ddots & \vdots \\
                  &        & I
            \end{pmatrix},
            \begin{pmatrix}
                I & \ldots & B_j    \\
                  & \ddots & \vdots \\
                  &        & I
            \end{pmatrix},
            \begin{pmatrix}
                I & \ldots & C_j    \\
                  & \ddots & \vdots \\
                  &        & I
            \end{pmatrix}
        \right) = 
    }
    \notag \\
    &\scalemath{0.6}{
        \begin{pmatrix}
            I &        &     \\
              & \ddots &  \\
              &        & A_J^{-1}
        \end{pmatrix}
        \left(
    	    \begin{pmatrix}
    	        I &        &     \\
    	           & \ddots &  \\
    	           &        & I
    	    \end{pmatrix},
            \begin{pmatrix}
                I & \ldots & I    \\
                  & \ddots & \vdots \\
                  &        & I
            \end{pmatrix},
            \begin{pmatrix}
                I & \ldots & B_j A_J^{-1}    \\
                  & \ddots & \vdots \\
                  &        & I
            \end{pmatrix},
            \begin{pmatrix}
                I & \ldots & C_j A_J^{-1}   \\
                  & \ddots & \vdots \\
                  &        & I
            \end{pmatrix}
        \right)
        \begin{pmatrix}
            I &        &     \\
              & \ddots &  \\
              &        & A_j
        \end{pmatrix}
    }
    \,. \label{eq:Z-factoring}
\end{align}
The common $S$ (left-hand-side) factor represents a Tezuka scambling, hence may be dropped.
Thus, without loss of generality, we may populate all blocks in the first row of the second-dimension (first of the three) matrix with identity matrix blocks.
Once we implement this last factoring, the exhaustive search reduces to this one-and-only quadruple of generator matrices:
\begin{equation}
    \setlength{\matrixBaselineShift}{-1.5pt}
    \left\{
    \binaryMatrix[\matrixBaselineShift]{32}{
        1............................... 
        .1.............................. 
        ..1............................. 
        ...1............................ 
        ....1........................... 
        .....1.......................... 
        ......1......................... 
        .......1........................ 
        ........1....................... 
        .........1...................... 
        ..........1..................... 
        ...........1.................... 
        ............1................... 
        .............1.................. 
        ..............1................. 
        ...............1................ 
        ................1............... 
        .................1.............. 
        ..................1............. 
        ...................1............ 
        ....................1........... 
        .....................1.......... 
        ......................1......... 
        .......................1........ 
        ........................1....... 
        .........................1...... 
        ..........................1..... 
        ...........................1.... 
        ............................1... 
        .............................1.. 
        ..............................1. 
        ...............................1 
    }\,,
    \binaryMatrix[\matrixBaselineShift]{32}{
        1.1.1.1.1.1.1.1.1.1.1.1.1.1.1.1. 
        .1.1.1.1.1.1.1.1.1.1.1.1.1.1.1.1 
        ..1...1...1...1...1...1...1...1. 
        ...1...1...1...1...1...1...1...1 
        ....1.1.....1.1.....1.1.....1.1. 
        .....1.1.....1.1.....1.1.....1.1 
        ......1.......1.......1.......1. 
        .......1.......1.......1.......1 
        ........1.1.1.1.........1.1.1.1. 
        .........1.1.1.1.........1.1.1.1 
        ..........1...1...........1...1. 
        ...........1...1...........1...1 
        ............1.1.............1.1. 
        .............1.1.............1.1 
        ..............1...............1. 
        ...............1...............1 
        ................1.1.1.1.1.1.1.1. 
        .................1.1.1.1.1.1.1.1 
        ..................1...1...1...1. 
        ...................1...1...1...1 
        ....................1.1.....1.1. 
        .....................1.1.....1.1 
        ......................1.......1. 
        .......................1.......1 
        ........................1.1.1.1. 
        .........................1.1.1.1 
        ..........................1...1. 
        ...........................1...1 
        ............................1.1. 
        .............................1.1 
        ..............................1. 
        ...............................1 
    }\,,
    \binaryMatrix[\matrixBaselineShift]{32}{
        1..1111..1111..1111..1111..1111. 
        .1111..1111..1111..1111..1111..1 
        ..1...11...1..1...11...1..1...11 
        ...1..1...11...1..1...11...1..1. 
        ....1..1.....111....111.....1..1 
        .....111....111.....1..1.....111 
        ......1........1......11......1. 
        .......1......11......1........1 
        ........1..1111.........111..111 
        .........1111..1........1..1111. 
        ..........1...11..........11...1 
        ...........1..1...........1...11 
        ............1..1............111. 
        .............111............1..1 
        ..............1...............11 
        ...............1..............1. 
        ................1..1111..1111..1 
        .................1111..1111..111 
        ..................1...11...1..1. 
        ...................1..1...11...1 
        ....................1..1.....111 
        .....................111....111. 
        ......................1........1 
        .......................1......11 
        ........................1..1111. 
        .........................1111..1 
        ..........................1...11 
        ...........................1..1. 
        ............................1..1 
        .............................111 
        ..............................1. 
        ...............................1 
    }\,,
    \binaryMatrix[\matrixBaselineShift]{32}{
        1.11.11.11.11.11.11.11.11.11.11.
        .11.11.11.11.11.11.11.11.11.11.1
        ..1....1..11..1....1..11..1....1
        ...1..11..1....1..11..1....1..11
        ....1.11....11.1.....11.....1.11
        .....11.....1.11....11.1.....11.
        ......1.......11.......1......1.
        .......1......1.......11.......1
        ........1.11.11..........11.11.1
        .........11.11.1........11.11.11
        ..........1....1...........1..11
        ...........1..11..........11..1.
        ............1.11.............11.
        .............11.............11.1
        ..............1................1
        ...............1..............11
        ................1.11.11.11.11.11
        .................11.11.11.11.11.
        ..................1....1..11..1.
        ...................1..11..1....1
        ....................1.11....11.1
        .....................11.....1.11
        ......................1.......11
        .......................1......1.
        ........................1.11.11.
        .........................11.11.1
        ..........................1....1
        ...........................1..11
        ............................1.11
        .............................11.
        ..............................1.
        ...............................1
    }
    \right\} \,,\label{eq:sz2 base matrices}
\end{equation}
here shown to 32-bit depth.
As per our exhaustive search, this quadruple represents the template of all binary-matrix-based (0, 4)-sequences, and all other possible variants may be derived by plugging back the degrees of freedom factored earlier, as we will discuss later.

Upon careful inspection of these matrices, the first important observation is that they are built exclusively from the $S$ block symbols in Eq.~\eqref{eq:S and Z symbols}, plus empty (0) blocks.
Further inspection, guided by a preliminary assumption that these sequences bear some relation to Faure's, reveals that these are actually Pascal matrices
\begin{equation}
    \mathcal{P}(a) = \left( p_{i,j}:\binom{j}{i} a^{j - i} \right)\,, \label{eq:pascal a}
\end{equation}
constructed from these four block symbols.

\section{SZ Sequences}
\label{sec:SZ sequences}

Based on our exhaustive search for binary $(0, 4)$-sequence generators and similar partial searches for $(0, 8)$ and $(0, 16)$ sequences, we find that what characterizes the $S$ set of $2\times 2$ block symbols, compared to the complementing $Z$ set in Eq.~\eqref{eq:S and Z symbols}, is that it is not only closed under addition, but also multiplication, and hence constitutes a finite field, as dictated by Wedderburn's little theorem, stating that ``every finite division ring is a field''.
In a nutshell, finite fields are sets of elements over which are defined analogous operations to addition and multiplication, and the set is closed under these two operations, with associated neutral elements analogous to 0 and 1, and with analogous definitions of additive and multiplicative inverses.
The only missing element in our alphabets to make a field is a zero, but in fact, this element has existed and been used all the way in our development: it is the symbol that is used to build the identity matrix in our set! This could be slightly confusing, since we have a ``grand'' identity matrix for the final set of generator matrices, built from the 0 symbol, and a ``Pascal-ed'' identity symbol used to build the second-dimension generator matrix.

Once we arrive at this conclusion for 4D, we may readily envision analogous constructions for higher power-of-2 dimensions, as summarized in Algorithm~\ref{alg:SZ sequence}.
\begin{algorithm} [tb]
    \SetKwInOut{KwIn}{Input}
    \SetKwInOut{KwOut}{Output}
    \caption{
        Constructing an SZ $(0, 2^q)$-sequence.
    }
    \label{alg:SZ sequence}
    \KwIn{(1) index $q$ of required power-of-2 dimension $s = 2^q$.}
    \KwOut{A set of $s$ binary generator matrices to produce a $(0, s)$-sequence.}
    Search for a set (alphabet) $\{x_i\}_{i=1}^{s-1}$ of $s-1$ invertible $q\times q$ matrices that is closed under addition, multiplication, and inversion\;
    Build a Pascal matrix $\mathcal{P}(x)$ of the desired numerical resolution for each symbol in the alphabet\;
    Along with an equally sized identity matrix, these constitute the desired set of generator matrices.
\end{algorithm}
Using this algorithm, we constructed SZ sequences for different values of $q$, and empirically verified their $(0, s)$-sequence property by building and validating all the sets of hybrid matrices.
It is worth noting here that, as a special case of our model, the conventional Pascal matrix in 2D Sobol/Faure constructions is the Pascal matrix of the identity of a 2D alphabet comprising granular 0 and 1 as their block matrices.
Thus, we hereinafter consider an all-zero $q\times q$ block matrix as an intrinsic symbol in our alphabets.

Thus, while we did not proactively choose to use fields, they automatically emerged, bringing us closer to Faure's sequences and their Niederreiter's extension.
Let us briefly abstract our empirical search results in Section~\ref{sec:exploration} to understand what happened:
We constructed the following four base-4 matrices:
\begin{equation}
    \scalemath{0.42}{
    \left\{\begin{array}{ll}
        \left(\begin{array}{cccccccccccccccc}
            1 & . & . & . & . & . & . & . & . & . & . & . & . & . & . & . \\
            . & 1 & . & . & . & . & . & . & . & . & . & . & . & . & . & . \\
            . & . & 1 & . & . & . & . & . & . & . & . & . & . & . & . & . \\
            . & . & . & 1 & . & . & . & . & . & . & . & . & . & . & . & . \\
            . & . & . & . & 1 & . & . & . & . & . & . & . & . & . & . & . \\
            . & . & . & . & . & 1 & . & . & . & . & . & . & . & . & . & . \\
            . & . & . & . & . & . & 1 & . & . & . & . & . & . & . & . & . \\
            . & . & . & . & . & . & . & 1 & . & . & . & . & . & . & . & . \\
            . & . & . & . & . & . & . & . & 1 & . & . & . & . & . & . & . \\
            . & . & . & . & . & . & . & . & . & 1 & . & . & . & . & . & . \\
            . & . & . & . & . & . & . & . & . & . & 1 & . & . & . & . & . \\
            . & . & . & . & . & . & . & . & . & . & . & 1 & . & . & . & . \\
            . & . & . & . & . & . & . & . & . & . & . & . & 1 & . & . & . \\
            . & . & . & . & . & . & . & . & . & . & . & . & . & 1 & . & . \\
            . & . & . & . & . & . & . & . & . & . & . & . & . & . & 1 & . \\
            . & . & . & . & . & . & . & . & . & . & . & . & . & . & . & 1
        \end{array}\right), &
        \left(\begin{array}{cccccccccccccccc}
            1 & 1 & 1 & 1 & 1 & 1 & 1 & 1 & 1 & 1 & 1 & 1 & 1 & 1 & 1 & 1 \\
            . & 1 & . & 1 & . & 1 & . & 1 & . & 1 & . & 1 & . & 1 & . & 1 \\
            . & . & 1 & 1 & . & . & 1 & 1 & . & . & 1 & 1 & . & . & 1 & 1 \\
            . & . & . & 1 & . & . & . & 1 & . & . & . & 1 & . & . & . & 1 \\
            . & . & . & . & 1 & 1 & 1 & 1 & . & . & . & . & 1 & 1 & 1 & 1 \\
            . & . & . & . & . & 1 & . & 1 & . & . & . & . & . & 1 & . & 1 \\
            . & . & . & . & . & . & 1 & 1 & . & . & . & . & . & . & 1 & 1 \\
            . & . & . & . & . & . & . & 1 & . & . & . & . & . & . & . & 1 \\
            . & . & . & . & . & . & . & . & 1 & 1 & 1 & 1 & 1 & 1 & 1 & 1 \\
            . & . & . & . & . & . & . & . & . & 1 & . & 1 & . & 1 & . & 1 \\
            . & . & . & . & . & . & . & . & . & . & 1 & 1 & . & . & 1 & 1 \\
            . & . & . & . & . & . & . & . & . & . & . & 1 & . & . & . & 1 \\
            . & . & . & . & . & . & . & . & . & . & . & . & 1 & 1 & 1 & 1 \\
            . & . & . & . & . & . & . & . & . & . & . & . & . & 1 & . & 1 \\
            . & . & . & . & . & . & . & . & . & . & . & . & . & . & 1 & 1 \\
            . & . & . & . & . & . & . & . & . & . & . & . & . & . & . & 1
        \end{array}\right),\\ \\
        \left(\begin{array}{cccccccccccccccc}
            1 & 2 & 3 & 1 & 2 & 3 & 1 & 2 & 3 & 1 & 2 & 3 & 1 & 2 & 3 & 1 \\
            . & 1 & . & 3 & . & 2 & . & 1 & . & 3 & . & 2 & . & 1 & . & 3 \\
            . & . & 1 & 2 & . & . & 2 & 3 & . & . & 3 & 1 & . & . & 1 & 2 \\
            . & . & . & 1 & . & . & . & 2 & . & . & . & 3 & . & . & . & 1 \\
            . & . & . & . & 1 & 2 & 3 & 1 & . & . & . & . & 3 & 1 & 2 & 3 \\
            . & . & . & . & . & 1 & . & 3 & . & . & . & . & . & 3 & . & 2 \\
            . & . & . & . & . & . & 1 & 2 & . & . & . & . & . & . & 3 & 1 \\
            . & . & . & . & . & . & . & 1 & . & . & . & . & . & . & . & 3 \\
            . & . & . & . & . & . & . & . & 1 & 2 & 3 & 1 & 2 & 3 & 1 & 2 \\
            . & . & . & . & . & . & . & . & . & 1 & . & 3 & . & 2 & . & 1 \\
            . & . & . & . & . & . & . & . & . & . & 1 & 2 & . & . & 2 & 3 \\
            . & . & . & . & . & . & . & . & . & . & . & 1 & . & . & . & 2 \\
            . & . & . & . & . & . & . & . & . & . & . & . & 1 & 2 & 3 & 1 \\
            . & . & . & . & . & . & . & . & . & . & . & . & . & 1 & . & 3 \\
            . & . & . & . & . & . & . & . & . & . & . & . & . & . & 1 & 2 \\
            . & . & . & . & . & . & . & . & . & . & . & . & . & . & . & 1
        \end{array}\right), &
        \left(\begin{array}{cccccccccccccccc}
            1 & 3 & 2 & 1 & 3 & 2 & 1 & 3 & 2 & 1 & 3 & 2 & 1 & 3 & 2 & 1 \\
            . & 1 & . & 2 & . & 3 & . & 1 & . & 2 & . & 3 & . & 1 & . & 2 \\
            . & . & 1 & 3 & . & . & 3 & 2 & . & . & 2 & 1 & . & . & 1 & 3 \\
            . & . & . & 1 & . & . & . & 3 & . & . & . & 2 & . & . & . & 1 \\
            . & . & . & . & 1 & 3 & 2 & 1 & . & . & . & . & 2 & 1 & 3 & 2 \\
            . & . & . & . & . & 1 & . & 2 & . & . & . & . & . & 2 & . & 3 \\
            . & . & . & . & . & . & 1 & 3 & . & . & . & . & . & . & 2 & 1 \\
            . & . & . & . & . & . & . & 1 & . & . & . & . & . & . & . & 2 \\
            . & . & . & . & . & . & . & . & 1 & 3 & 2 & 1 & 3 & 2 & 1 & 3 \\
            . & . & . & . & . & . & . & . & . & 1 & . & 2 & . & 3 & . & 1 \\
            . & . & . & . & . & . & . & . & . & . & 1 & 3 & . & . & 3 & 2 \\
            . & . & . & . & . & . & . & . & . & . & . & 1 & . & . & . & 3 \\
            . & . & . & . & . & . & . & . & . & . & . & . & 1 & 3 & 2 & 1 \\
            . & . & . & . & . & . & . & . & . & . & . & . & . & 1 & . & 2 \\
            . & . & . & . & . & . & . & . & . & . & . & . & . & . & 1 & 3 \\
            . & . & . & . & . & . & . & . & . & . & . & . & . & . & . & 1
        \end{array}\right)
    \end{array}\right\}
    }\,.
\end{equation}
These are exactly the ones in Eq.~\eqref{eq:sz2 base matrices}, just encoding each distinct $2\times 2$ block as a base-4 digit, namely indexed after 0 and the three $S$ symbols in Eq.~\eqref{eq:S and Z symbols}.
When multiplied by a base-4 column vector representation of a sequence number, these specific matrices generate a (0, 4)-sequence in base 4, i.e., a sequence of 4D points that is fully multi-stratified for powers-of-4 number of points.
However, multiplication is not the standard arithmetic modulo-4 one, but uses a special table
\begin{equation}
\begin{array}{cc|cccc}
      & Z & 0 & 1 & 2 & 3 \\
    S &   &   &   &   &  \\\hline
    0 &   & 0 & 0 & 0 & 0 \\
    1 &   & 0 & 1 & 2 & 3 \\
    2 &   & 0 & 2 & 3 & 1 \\
    3 &   & 0 & 3 & 1 & 2
\end{array}\,,
\end{equation}
that is built to amend the degenerate multiplication table modulo 4, preparing it for Faure construction.
Here $S$ refers to the left multiplicand digit from the matrix and $Z$ to the right one from the vector.
Finally,
\emph{this arithmetic table is not built arbitrarily, nor looked up, but is intrinsic to the matrix-block alphabet, and is evaluated automatically via GF(2) matrix-vector multiplication:}
\begin{equation}
    \settoheight{\matrixEntryWidth}{X}
    \setlength{\matrixEntryWidth}{0.555\matrixEntryWidth}
    \begin{pmatrix}
        \binaryMatrix[-0.7]{2}{ .. .. } \\
        \binaryMatrix[-0.7]{2}{ 1. .1 } \\
        \binaryMatrix[-0.7]{2}{ .1 11 } \\
        \binaryMatrix[-0.7]{2}{ 11 1. }
    \end{pmatrix}
    \begin{pmatrix}
        \binaryMatrix[-1]{1}{..} &
        \binaryMatrix[-1]{1}{1.} &
        \binaryMatrix[-1]{1}{.1} &
        \binaryMatrix[-1]{1}{11}
    \end{pmatrix}
    =
    \begin{pmatrix}
        \binaryMatrix[-1]{1}{..} & \binaryMatrix[-1pt]{1}{..} & \binaryMatrix[-1pt]{1}{..} & \binaryMatrix[-1pt]{1}{..} \\
        \binaryMatrix[-1]{1}{..} & \binaryMatrix[-1pt]{1}{1.} & \binaryMatrix[-1pt]{1}{.1} & \binaryMatrix[-1pt]{1}{11} \\
        \binaryMatrix[-1]{1}{..} & \binaryMatrix[-1pt]{1}{.1} & \binaryMatrix[-1pt]{1}{11} & \binaryMatrix[-1pt]{1}{1.} \\
        \binaryMatrix[-1]{1}{..} & \binaryMatrix[-1pt]{1}{11} & \binaryMatrix[-1pt]{1}{1.} & \binaryMatrix[-1pt]{1}{.1}
    \end{pmatrix}\,.    
\end{equation}
\settoheight{\matrixEntryWidth}{X}%
\setlength{\matrixEntryWidth}{0.555\matrixEntryWidth}%
Please mind the bit-reversed ordering: top row is least significant bit, so \binaryMatrix[-1.4]{1}{1.} is 1 and \binaryMatrix[-1.4]{1}{.1} is 2.

Even though we have arrived at this construction experimentally and independently, as detailed in the preceding Section~\ref{sec:exploration}, once abstracted this way, we may see a close similarity to the analytically developed construction of Niederreiter \shortcite{Niederreiter87Point,Niederreiter92Random} that extends Faure's sequences to power-of-prime bases. Further, the uniqueness of finite fields makes us strongly believe that our construction does actually coincide with Niederrieter's.
This remains a hypothesis, though, awaiting for analytical validation, since the multiple degrees of freedom in both constructions makes them produce different realizations, which makes empirical verification difficult.
In all cases, both constructions span the same universe of $(0, 2^q)$-sequences in base $2^q$.
It is the emphasized text in the preceding description, however, that is the essential difference between our construction and Niederrieter's: instead of using symbolic GF($2^q$) arithmetic with lookup tables, as in the standard implementation of Niederreiter sequences \cite{Bratley1992Implementation}, our field elements are embedded as block matrices in GF(2) generator matrices that  handle both the field arithmetic and the bijections operators in Niederreiter's construction.
This makes our binary-based construction far more efficient and scalable.
Another important advantage is that our experimental approach lead us to discover nesting and ensembling, which makes a significant difference in graphics applications, as we will see later in Section~\ref{sec:rendering}.

\subsection{Alphabets}
\label{sec:alphabets}

Identifying alphabets as finite fields brings with many insights and implications.
For example, it implies commutativity of matrix multiplication over the set, which we will use later for randomization.
The most relevant property for our construction, however, is that in fields like our alphabets there exists a symbol, in fact many, that multiplicatively cycles through all other non-zero symbols; that is, it can generate all the non-zero elements as its powers.
Such a symbol is called a ``primitive'' and commonly designated as $\alpha$.
Then the alphabet may be written as
\begin{equation}
    \Sigma = \{0, I, \alpha, \alpha^2, \ldots, \alpha^{q-2}\}\,.
\end{equation}

Aside from analytical insights and proofs, alpha elements prove quite helpful in practice.
The first facility they can offer is a straightforward search plan for alphabets, as summarized in Algorithm~\ref{alg:alpha-search}, replacing complex and time-consuming stack-based searches we implemented in our exploration phase.
\begin{algorithm} [tb]
    \SetKwInOut{KwIn}{Input}
    \SetKwInOut{KwOut}{Output}
    \caption{
        AlphaSearch: Searching for an alphabets.
    }
    \label{alg:alpha-search}
    \KwIn{(1) A power of 2 size of elements $s = 2^q$,\\
          (2) A timeout number of attempts T.}
    \KwOut{An alphabet of $s$ block matrices.}
    \For{$i \gets 0$ \KwTo $T - 1$} {
        Construct a random invertible $q\times q$ matrix block $x$\;
        Repeatedly multiply $x$ by itself until you get $I$, and record the needed number of multiplications: its ``root index'' $n$\;
        \If{ $n = s - 1$ } {
            \Return{$\left\{ 0, \left\{x^i\right\}_{i=0}^{s - 2} \right\}$}\;
        }
    }
    Return a timeout message.
\end{algorithm}
Secondly, noting that an alpha element can belong to a unique alphabet, the search algorithm can easily be modified to enumerate all alphabets, dividing the number of alpha elements in the universe of invertible matrices by the number of alphas in a single alphabet.
We performed such a search over the feasible range, and obtained
\begin{equation}
    \left(\left| \left\{ \Sigma(q)\right\}_{q=1}^{6} \right|\right)
    = (1, 1, 8, 336, 64512, 53329920)\,.
\end{equation}
A brief Internet search identified this with OEISA258745~\shortcite{OEISA258745}, described as ``Order of general affine group AGL(n,2) (=A028365(n)) divided by (n+1)''.
This gives a hint for the possibility of building alphabets algorithmically, which we leave for future research.

\subsection{Randomization}

With the template set of generator matrices built up exclusively of alphabet building blocks, we may now randomize the set by plugging back the degrees of freedom we factored earlier.
The extension rows we removed in Eq.~\eqref{eq:R AInv C + X}, along with the $X$ diagonal blocks we factored later, exactly identify with the Tezuka's~\shortcite{Tezuka1994Generalization} scrambling mentioned in Section~\ref{sec:scrambling and shuffling}.
Beyond that, our model readily supports Owen's~\shortcite{Owen95Randomly} scrambling and Faure-Tezuka~\shortcite{Faure02Another} shuffling in both $2$ and $2^q$ bases; as well as Xor-scrambling~\cite{Kollig02Efficient}.

\subsection{Nesting}

The key idea of nesting, introduced in the exploration phase, is to find an alphabet over $2^{2q}$ that embeds in its symbols the symbols of a given alphabet over $2^q$, in such a way that the generators of the nested sequence may be reproduced using sequence-preserving manipulations of the corresponding generators of the nesting sequence.
Toward that end, we start by extracting symbols of the nesting alphabet from generators of the nested sequence.
We note that 2$\times$2 blocks of the nested generator tile a single block of the nesting generator.
The symbol of the nesting alphabet is found in the second block of the first row of the Pascal matrix, but we first have to factor out the first block, leaving only an identity;
hence
\begin{align}
    \langle a \rangle
    =
    \begin{pmatrix} I & a \\   & I \end{pmatrix}^{-1}
    \begin{pmatrix} a^2 & a^3 \\   & a^2 \end{pmatrix}
    = \begin{pmatrix} a^2 &  \\   & a^2 \end{pmatrix}\,, \label{eq:<a>}
\end{align}
where $\langle a \rangle$ denotes a symbol of the nesting alphabet embedding symbol $a$ of the nested sequence alphabet.
We then search for a $2^{2q}$ alphabet that contains this subset.
A brute-force approach to such a search is far more difficult than before, since we seek quite specific alphabets.
We therefore generated and analyzed all nesting alphabets in a feasible range, and luckily managed to find a pattern that all nesting alphabets are built exclusively from the nested alphabet symbols.
This reduces the search space dramatically from $2^{4q^2}$ to $2^{4q}$, which makes it feasible to reach 64K dimensions.
This reduction not only enables us to find working alphabets, but to uniformly sample them.

Once the alphabet is found, the generator matrices are built as per Algorithm~\ref{alg:SZ sequence}, then the generator matrices of embedded symbols are multiplied by a factor
\begin{equation}
     \mathcal{P} (a) = \begin{pmatrix} I & a \\   & I \end{pmatrix} \mathcal{P} (\langle a \rangle)\,, \label{eq:nested generator restoration}
\end{equation}
to restore the original matrices, which provably works by expanding the multiplication by the block entry $(j, i)$
\begin{align}
    \scalemath{0.85}{
     \begin{pmatrix} I & a \\   & I \end{pmatrix}
     \begin{pmatrix} \langle a \rangle^{j - i} \end{pmatrix}
     =
     \begin{pmatrix} I & a \\   & I \end{pmatrix}
     \begin{pmatrix} a^{2j - 2i} & \\ & a^{2j - 2i} \end{pmatrix}
     = \begin{pmatrix} a^{2j - 2i} & a^{(2j + 1) - 2i} \\ & a^{2j - 2i} \end{pmatrix}}\,,
\end{align}
which match the entries in $\mathcal{P}(a)$, then the coefficients
\begin{equation}
     \begin{pmatrix}
         \binom{2j}{2i    } & \binom{2j + 1}{2i    } \\ \\
         \binom{2j}{2i + 1} & \binom{2j + 1}{2i + 1} 
     \end{pmatrix}\,,
\end{equation}
are matched using  Lucas' Theorem.
Note that for the nesting set, this multiplication represents a Tezuka scrambling, hence preserves the structure of the sequence.

We conclude this section by making a remark that nesting, as described above, works only by doubling $q$, i.e., squaring the dimension in each upsampling step of the sequence space, not by doubling the dimension as we would hope.

\subsection{Ensembling}

Ensembling is less obvious than nesting, since it depends on a non-obvious property.
We start with a nested sequence, and the key idea is to try to make subsequent bands of dimensions of the nesting sequence, with respect to each other, look identical to their counterparts in the nested band.
This starts by arranging the symbols in a suitable order, as outlined in Algorithm~\ref{alg:nestsort},
\begin{algorithm} [tb]
    \SetKwInOut{KwIn}{Input}
    \SetKwInOut{KwOut}{Output}
    \caption{
        NestSort: order alphabets to enable ensembling.
    }
    \label{alg:nestsort}
    \KwIn{An alphabet $\Sigma$ of $s = 2^q$ block-matrix symbols.}
    \KwOut{An ordered alphabet with a nested substructure.}
    Ensure the placement of the 0 and $I$ symbols in the first and second slots, respectively\;
    $k \gets 2$\;
    \While{$k < s$}{
        \For{$i \gets 1$ \KwTo $k-1$}{
            $x \gets \Sigma[i] + \Sigma[k]$\;
            Locate the slot $j$ containing $x$\;
            Swap $\Sigma[i+k]$ with $\Sigma[j]$\;
        }
        $k \gets 2\cdot k$\;
    }
\end{algorithm}
which arranges the symbols in a literally nested algebraic structure of bands
\begin{align}
    B_1 \leftarrow (0, I)\,,\;
    B_2 \leftarrow (B_1, B_1 + x_1)\,,\;
    B_3 \leftarrow (B_2, B_2 + x_2)\,,\; \ldots \label{eq:nesting}
\end{align}
This idea is based on an interesting property
\begin{equation}
    \mathcal{P}(a + x) = \mathcal{P}(a)\mathcal{P}(x) \label{eq:Pascal sun-to-produt}\,,
\end{equation}
of Pascal matrices that summing the symbols is equivalent to multiplying the corresponding Pascal matrices.
A sketch of a proof starts by noting that each entry
\begin{equation}
    \binom{j}{i} (a + x)^{j - i}\,,
\end{equation}
of $\mathcal{P}(a + x)$ on the right-hand side embodies a binomial expansion with the same range of powers as the corresponding row-column product on the left-hand side, and it then remains to equate the binomial coefficients.
Applying this property to the nested structure in Eq.~\eqref{eq:nesting}, we note that adding a symbol to each subset is equivalent to multiplying its generator matrix by those of the subset, with a net effect of reordering the points in the subset while preserving their geometry.
That is, the sub-sequence of each band is just a reordering of the sequence of the reference band, hence can be turned into a nested sequence by the same restoration in Eq.~\eqref{eq:nested generator restoration}, actually the identical sequence, but in a different order.
This can be observed in Fig.~\ref{fig:teaser}(b), where all (0, 2) components look identical.

Ensembling, as described above, imposes substantial correlation between corresponding dimensions in different bands, which might cause undesirable effects in some applications.
To mitigate this, we may shuffle the points in different bands, as done when pairing independent 2D sequences~\cite{Burley2020Scrambling}, but that would break the high-dimensional structure in our case. Fortunately, we may take advantage of the commutative multiplication property of alphabets mentioned in Section~\ref{sec:alphabets}: if we multiply any of the Pascal generator matrices from left by any symbol from the alphabet, it is multiplied by all blocks, then the multiplication order can be exchanged, and the symbol can be factored on the right.
The operation on the left is a Tezuka \emph{scrambling}, hence preserves the structure of the sequence, but it is effectively equivalent to a Faure--Tezuka \emph{shuffling} (see Section~\ref{sec:scrambling and shuffling}) that reorders the coordinates in the concerned dimension.
To use this idea with ensembling, we use 4D symbols to shuffle 2D bands, 16D symbols to shuffle 4D bands, and so on. More details are in our supplementary code.

\section{Results and Applications}
\label{sec:results}

To evaluate the performance of SZ-sequences, we have evaluated them both for the numerical integration of synthetic functions
as well as for rendering.
We further include empirical measurements of discrepancy as well as analysis of their power spectra.   
Our implementation is included in the supplemental material.

\begin{figure*}[tb]
  \centering
    \begin{tabular}{c@{\hspace{0.1cm}}c@{\hspace{0.1cm}}c@{\hspace{0.1cm}}c@{\hspace{0.1cm}}c}

    \raisebox{-1.6cm}{\rotatebox[origin=r]{90}{Linear ($g^1$)}} &
    \begin{subfigure}[t]{0.22\linewidth}
      \centering
     \caption*{\hspace{0.7cm}$f^D_{1,2} * f^D_{3,4}$}
     \vbox{
            \includegraphics[height=3.5cm]{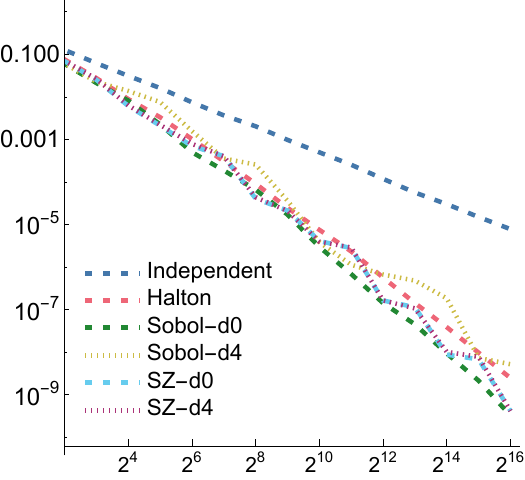}
        }
    \end{subfigure} &
    \begin{subfigure}[t]{0.22\linewidth}
      \centering
      \caption*{\hspace{0.5cm}$f^D_{1,2,3,4}$}
      \vbox{
            \includegraphics[height=3.5cm]{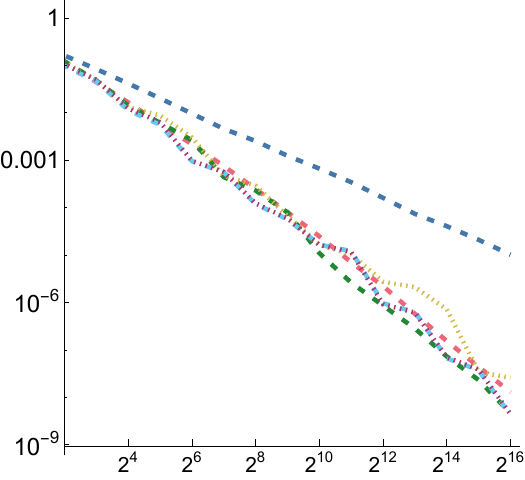}
        }
    \end{subfigure} &
    \begin{subfigure}[t]{0.22\linewidth}
      \centering
       \caption*{\hspace{0.5cm}$f^D_{1,2} * f^D_{3,4} + f^D_{5,6} * f^D_{7,8}$}
       \vbox{
            \includegraphics[height=3.5cm]{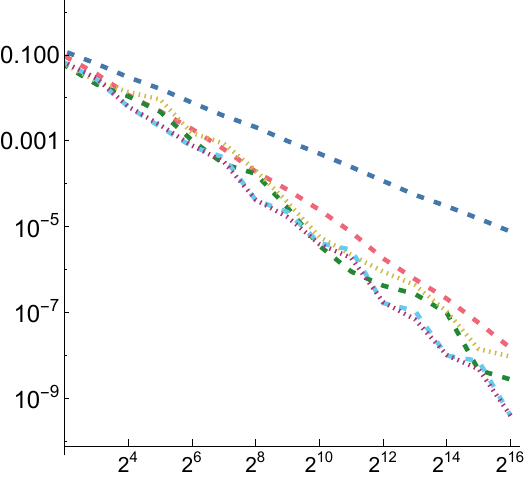}
        }
    \end{subfigure} &
    \begin{subfigure}[t]{0.22\linewidth}
      \centering
      \caption*{\hspace{0.7cm}$f^D_{1,2} * f^D_{1,3} * f^D_{1,4} * f^D_{2,3} * f^D_{2,4} * f^D_{3,4}$}
        \vbox{
            \includegraphics[height=3.5cm]{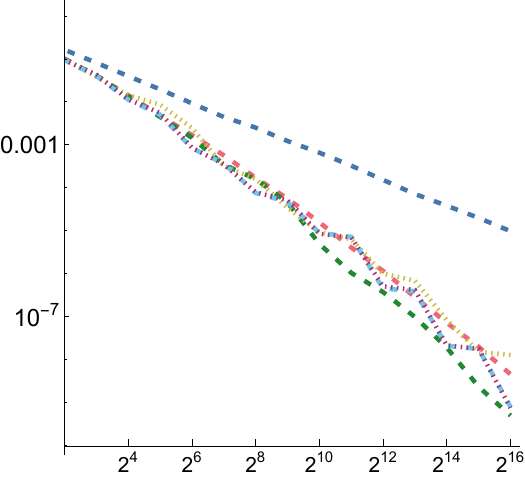}
        }
    \end{subfigure} \\

    \raisebox{2.7cm}{\rotatebox[origin=r]{90}{Gaussian ($g^\infty$)}} &
    \begin{subfigure}[t]{0.22\linewidth}
      \centering
        \vbox{
            \includegraphics[height=3.5cm]{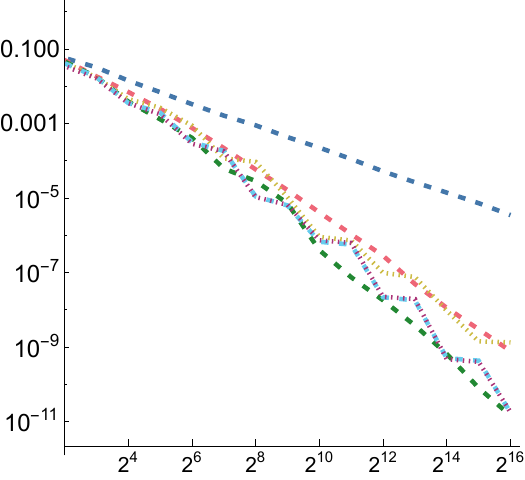}
        }
    \end{subfigure} &
    \begin{subfigure}[t]{0.22\linewidth}
      \centering
        \vbox{
            \includegraphics[height=3.5cm]{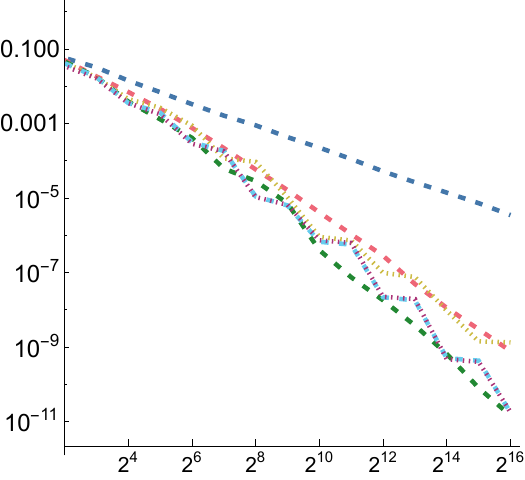}
        }
    \end{subfigure} &
    \begin{subfigure}[t]{0.22\linewidth}
      \centering
        \vbox{
            \includegraphics[height=3.5cm]{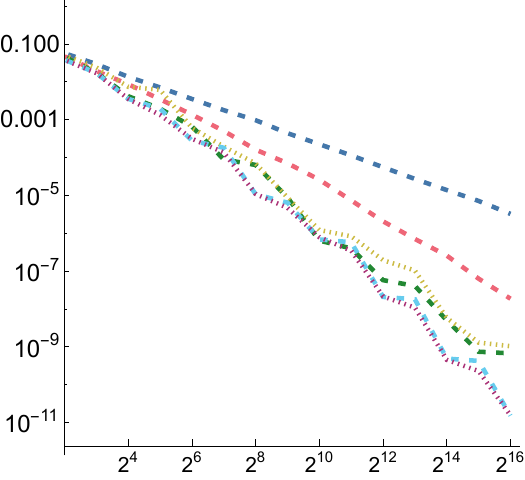}
        }
    \end{subfigure} &
    \begin{subfigure}[t]{0.22\linewidth}
      \centering
        \vbox{
            \includegraphics[height=3.5cm]{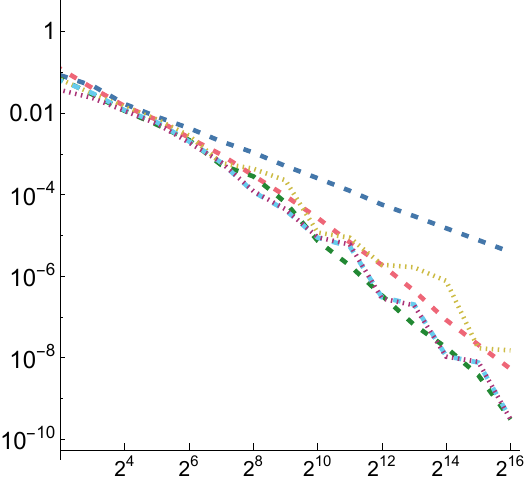}
        }
    \end{subfigure} \\

    \end{tabular}
    \caption{Mean relative squared error when integrating of a variety of analytic 4- and 8-dimensional functions.
      In these plots, ``Sobol-d0'' and ``Sobol-d4'' denote Sobol points starting at dimension 0 and 4, respectively,
      and similarly for SZ.
      SZ-0 performs as well as Sobol at power-of-4 sample counts, though is sometimes worse at intermediate power-of-2 counts.
      In some cases (e.g., the third column of $g^\infty$), it gives significantly lower error.
      SZ-4 generally performs as well as SZ-0, while Sobol-4 often has higher error than Sobol-0.
      See Fig.~\ref{fig:g0-analytic-function-error} for results with the $g^0$ function.
    }
    \label{fig:analytic-function-error}
\end{figure*}


\subsection{Numerical Integration}
\label{sec:integration}

In order to compare the performance of SZ points to other point sets for integration,
we measured mean relative squared error (MRSE) when integrating a variety of simple analytic
functions.
We follow the approach developed by Jarosz et al.~\shortcite[Section 5.2]{Jarosz19Orthogonal},
which we summarize here.
Three 1D functions are used as building blocks
\begin{align}
  g^0(r) &= 1 - \mathrm{binaryStep}(r, r_\mathrm{end})\,, \\
  g^1(r) &= 1 - \mathrm{linearStep}(r, r_\mathrm{start}, r_\mathrm{end})\,, \\
  g^\infty(r) &= \mathrm{e}^{-r^2/(2\sigma^2)}\,,
\end{align}
with $r_\mathrm{end}=3/\pi$, $r_\mathrm{start}=r_\mathrm{end}-0.2$, and $\sigma=1/3$.
Projection of an $n$-dimensional point $p$ to a set of dimensions $d_i$ is denoted by
\begin{equation}
  p_{d_1,\ldots,d_n} = ( p_{d_1}, \ldots, p_{d_n} )\,,
\end{equation}
$n$-dimensional functions are then defined by taking the vector norm of projected points
\begin{equation}
  f_{d_1,\ldots,d_n}^D(p) = g^D(\| p_{d_1}, \ldots, p_{d_n} \|).
\end{equation}
These are plots of associated 2D functions:
\begin{figure}[h]
  \begin{tabular}{ccc}
    \begin{subfigure}[t]{0.2\columnwidth}
      \centering
      \includegraphics[width=0.7\columnwidth]{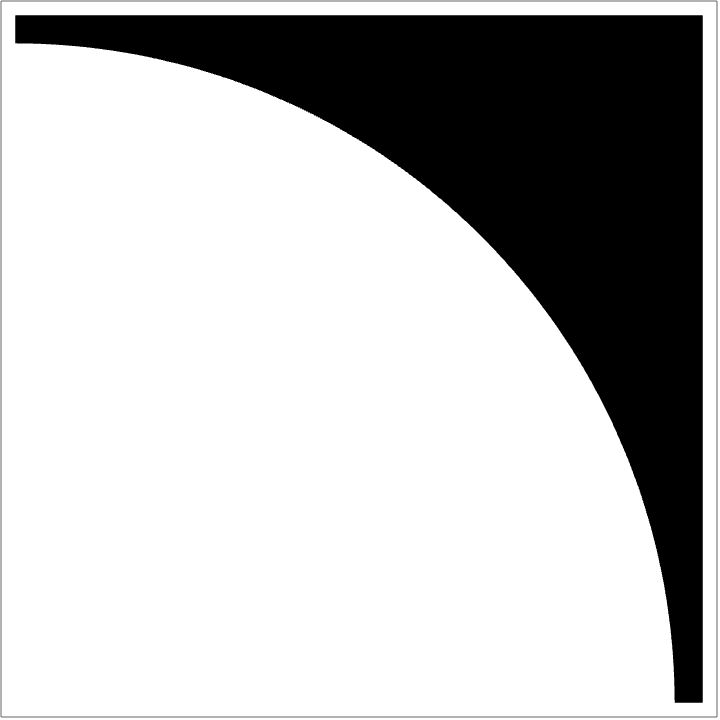}
      \caption*{$f_{1,2}^0$}
    \end{subfigure}&
    \begin{subfigure}[t]{0.2\columnwidth}
      \centering
      \includegraphics[width=0.7\columnwidth]{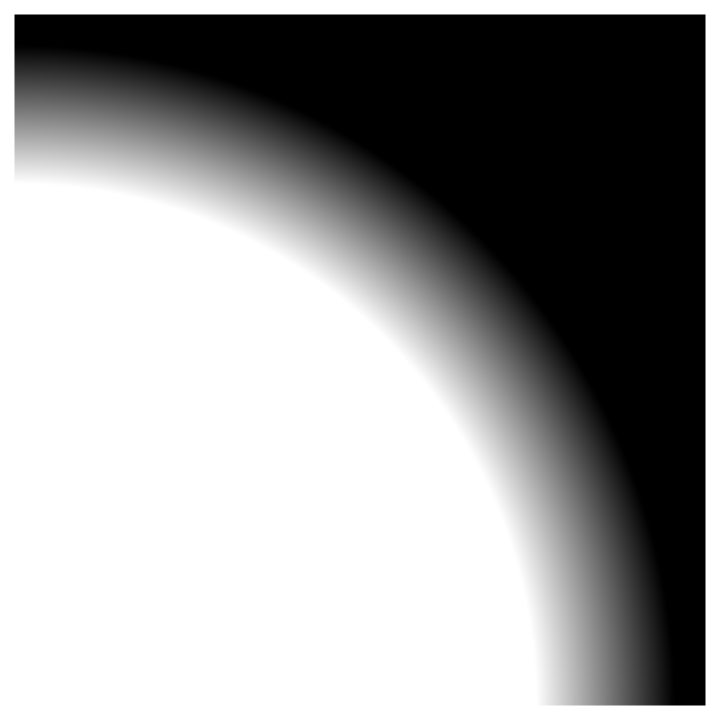}
      \caption*{$f_{1,2}^1$}
    \end{subfigure}&
    \begin{subfigure}[t]{0.2\columnwidth}
      \centering
      \includegraphics[width=0.7\columnwidth]{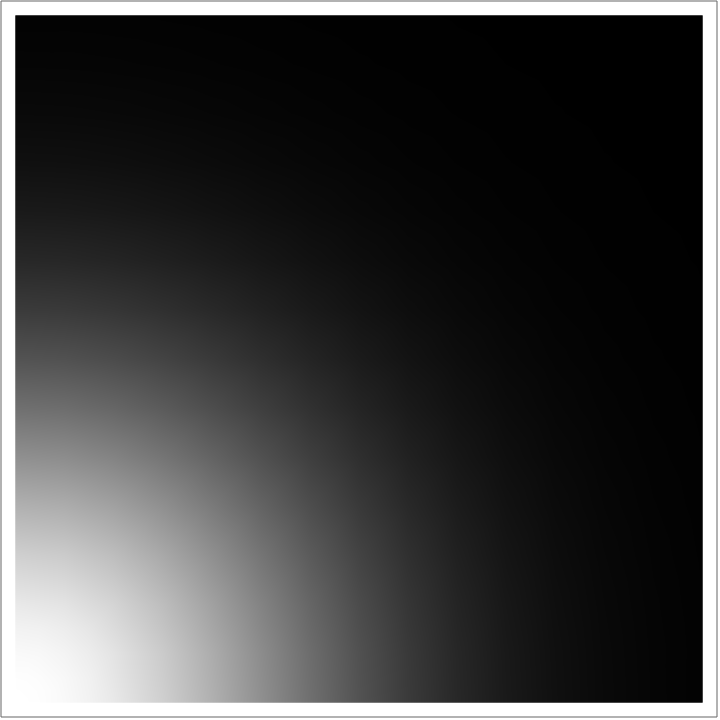}
      \caption*{$f_{1,2}^\infty$}
    \end{subfigure}\\
  \end{tabular}
\end{figure}

We used four forms of functions for our tests:
\begin{itemize}
\item $f^D_{1,2} * f^D_{3,4}$: the product of two 2D functions.
\item $f^D_{1,2,3,4}$: a fully 4D function.
\item $f^D_{1,2} * f^D_{3,4} + f^D_{5,6} * f^D_{7,8}$: the sum of the product of two 2D functions.
\item $f^D_{1,2} * f^D_{1,3} * f^D_{1,4} * f^D_{2,3} * f^D_{2,4} * f^D_{3,4}$: the product of functions of all 2D projections of a 4D point.
\end{itemize}
For each of these and for each of the three $g$ functions defined above, we computed a reference value and then
ran 1,024 independent trials, each taking up to $2^{16}$ samples.
Owen scrambling was used to randomize the low-discrepancy sequences.
In addition to our SZ sequence and the Sobol sequence, we also measured results with independent uniform random numbers
and with the Halton sequence.
For SZ and Sobol, we also measured results using samples starting at the 4th dimension,
to reflect the case in rendering where such integrands might be encountered after a few dimensions of the sequence had already been consumed.
We will use ``SZ-d0'' and ``SZ-d4'' to distinguish these, and similarly for Sobol.
We measured MRSE at all power-of-two number of samples

\begin{figure*}
    \setlength{\unit}{(\textwidth - 5\gap)/6}                        
    {\scriptsize \centering
    \begin{tabular*}{1\textwidth}{@{}c@{\extracolsep{\fill}}c@{\extracolsep{\fill}}c@{\extracolsep{\fill}}c@{\extracolsep{\fill}}c@{\extracolsep{\fill}}c@{}}
        \includegraphics[width=1\unit]{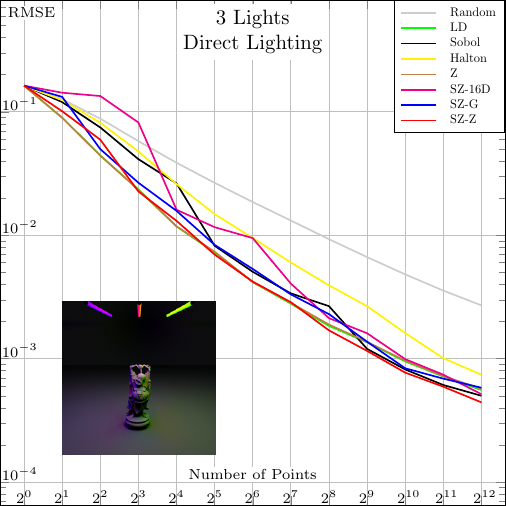}&%
        \includegraphics[width=1\unit]{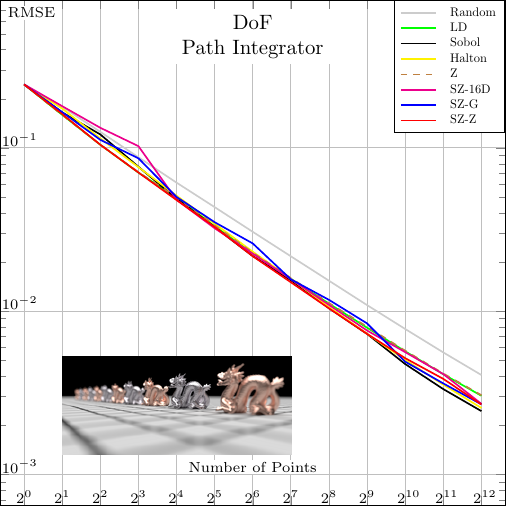}&%
        \includegraphics[width=1\unit]{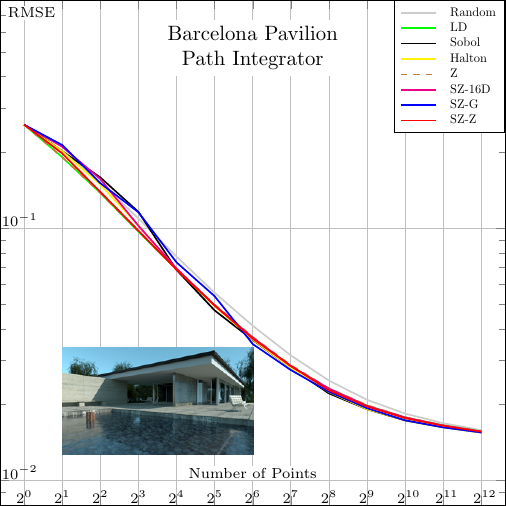}&%
        \includegraphics[width=1\unit]{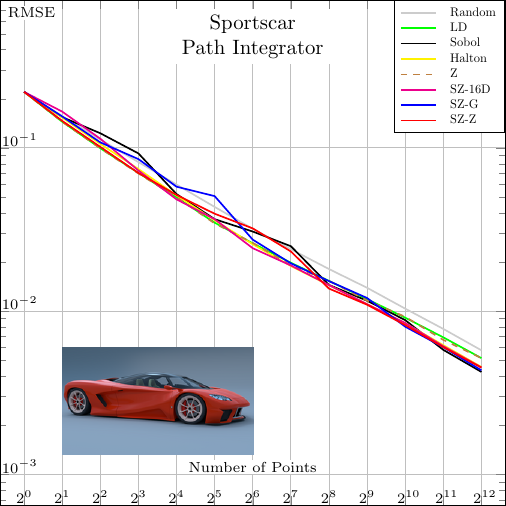}&%
        \includegraphics[width=1\unit]{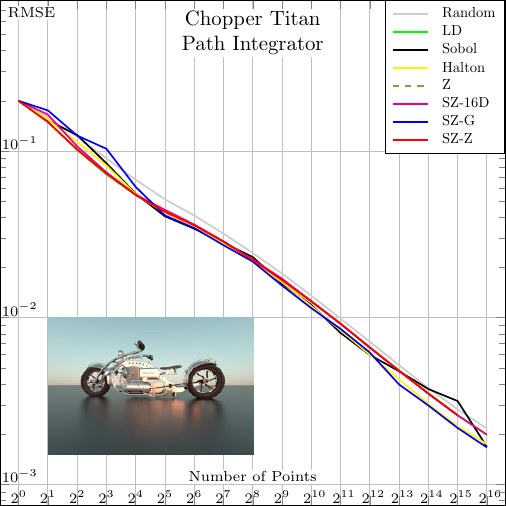}&%
        \includegraphics[width=1\unit]{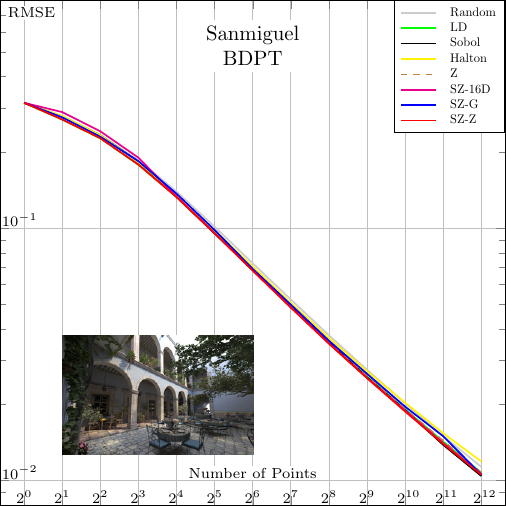}\\[1mm]%
        (a) Complex Lighting & (b) Depth of Field & (c) Complex Light Transport & (d) Environment Map & (e) Glossy Surfaces & (f) Bidirectional Path Tracing
    \end{tabular*}
    }
    \caption{
        Plots comparing RMSE of SZ-based samplers to state-of-the-art samplers for various common test scenes (insets).
        G/Z refer to global/Z indexing strategy \protect{\cite{Pharr23PBRT}}, and SZ-16D is a raw (non-nested) 16D SZ sequence.
        Beneath each plot we highlight an interesting feature of the scene.
        The range is extended in (e) as an example of the complex convergence behaviors.
    }
    \label{fig:rmse}
\end{figure*}

For all functions, performance of all samplers other than independent is similar for $g^0$, so we will focus on $g^1$ and $g^\infty$
in the following.
Error plots for $g^1$ and $g^\infty$ are shown in Fig.~\ref{fig:analytic-function-error}; the plot for $g^0$ can be seen in Fig.~\ref{fig:g0-analytic-function-error} in our supplemental.
For both the first and second function forms, SZ-d0 matches Sobol-d0's performance at power-of-4 sample counts,
though it has higher error at intermediate powers of two.
However, SZ-d4 delivers equally good error to SZ-d0, while Sobol-d4 is significantly worse.
We would expect this trend to continue at higher dimensions, as SZ maintains its $(0,2)$- and $(0,4)$-sequences,
while the quality of lower-dimensional projections of Sobol points worsens.
For the third function form, the sum of the product of two 2D functions, both SZ-d0 and SZ-d4 significantly outperform the Sobol sampler.
We believe that this is due both to it providing $(0,2)$-sequences for each of the constituent 2D functions
but also proving $(0,4)$-sequences for each term.
Finally, in the product of all of the 2D projections of a 4D point,we again see SZ-d0 and SZ-d4 matching Sobol at power-of-4 sample counts,
and SZ-d4 having much lower error than Sobol-d4. 

\subsection{Rendering}
\label{sec:rendering}

To evaluate the effectiveness of SZ sequences for rendering, we first implemented a \texttt{Sampler} in \emph{pbrt}~\cite{Pharr16PBRT},
using the global sampling strategy, where the first two dimensions are used for the image plane, and another using Z-Sampling~\cite{Ahmed2020Screen}, where a shuffled Z-sequence of the pixels is used to index the samples.
We modified the later, however, to sample all dimensions at once using SZ sequences.

In Fig.~\ref{fig:rmse}, we compare these SZ-based samplers to state-of-the-art samplers for various common test scenes.
We also included a raw (non-nested) 16D Z-indexed SZ sampler to give a rough idea about the expected performance of a randomly taken Niederreiter's (0, 16) and Faure's (0, 17) sequences.
The plots suggest that the baseline performance of SZ is generally superior to Halton, Faure, and Niederreiter, thanks to finer granularity, and is in the same range as Sobol.
We also note that the Z-indexed SZ sampler offers a decent aliasing/noise reduction compromise at low/high sampling rates, respectively.
That is, it outperforms global samplers at low sampling rates and the original Z sampler at higher rates.

\begin{figure*}[tb]
  \centering
    \begin{tabular}{c@{\hspace{0.01cm}}c@{\hspace{0.01cm}}c@{\hspace{0.01cm}}c@{\hspace{0.01cm}}c@{\hspace{0.01cm}}c@{\hspace{0.01cm}}c}

    \begin{subfigure}[t]{0.3\linewidth}
      \centering
        \caption*{\textsc{Reference}}
        \vbox{
            \includegraphics[height=2.8cm]{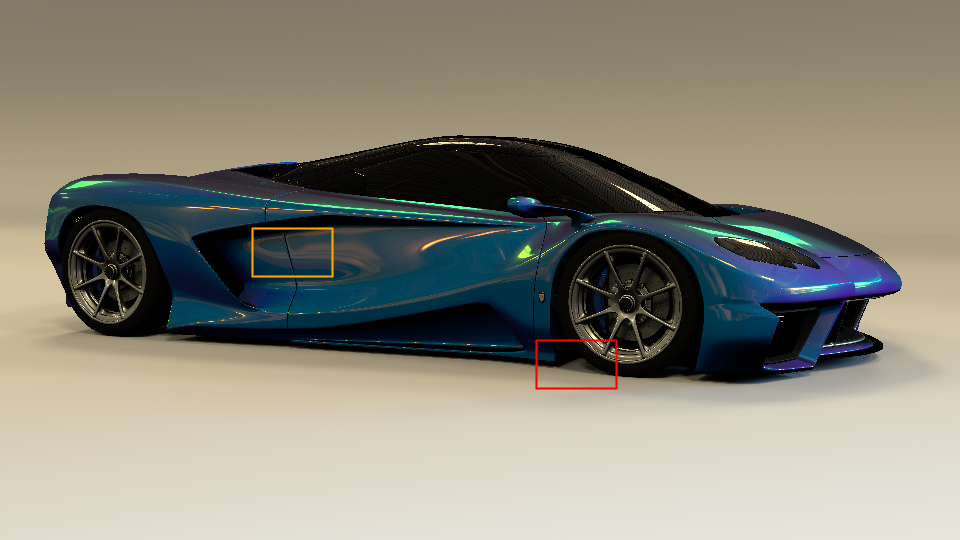}
        }
        \caption*{MRSE / \FLIP\\~}
    \end{subfigure} &
    \begin{subfigure}[t]{0.13\linewidth}
      \centering
        \caption*{\textsc{Sobol}}
        \vbox{
            \includegraphics[height=1.3cm]{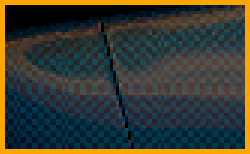}\vspace{0.2cm}
            \includegraphics[height=1.3cm]{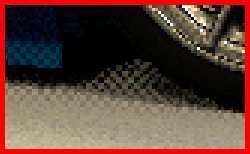}
        }
        \caption*{0.037195 / 0.049664\\~}
    \end{subfigure} &
    \begin{subfigure}[t]{0.13\linewidth}
      \centering
        \caption*{\textsc{Sobol MRSE}}
        \vbox{
            \includegraphics[height=1.3cm]{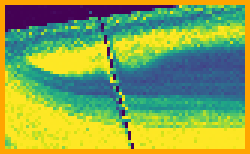}\vspace{0.2cm}
            \includegraphics[height=1.3cm]{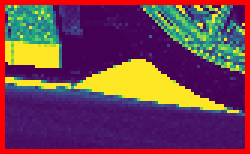}
        }
    \end{subfigure} &
    \begin{subfigure}[t]{0.13\linewidth}
      \centering
        \caption*{\textsc{SZ}}
        \vbox{
            \includegraphics[height=1.3cm]{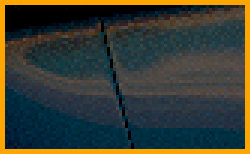}\vspace{0.2cm}
            \includegraphics[height=1.3cm]{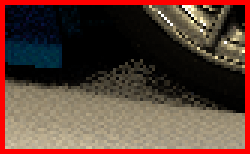}
        }
        \caption*{\textbf{0.019270} / \textbf{0.047296}\\~}
    \end{subfigure} &
    \begin{subfigure}[t]{0.13\linewidth}
      \centering
        \caption*{\textsc{SZ MRSE}}
        \vbox{
            \includegraphics[height=1.3cm]{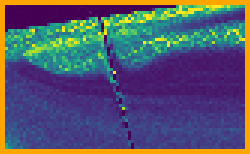}\vspace{0.2cm}
            \includegraphics[height=1.3cm]{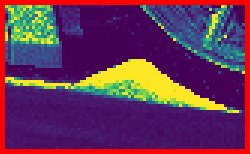}
        }
    \end{subfigure} &
    \begin{subfigure}[t]{0.13\linewidth}
        \centering
        \caption*{\textsc{Reference}}
        \vbox{
            \includegraphics[height=1.3cm]{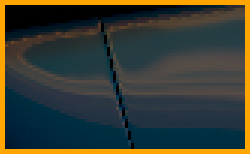}\vspace{0.2cm}
            \includegraphics[height=1.3cm]{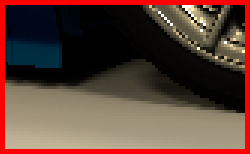}
        }
    \end{subfigure} &

    \begin{subfigure}[t]{0.03\linewidth}
      \raisebox{-1.125\height}{%
        \begin{tikzpicture}
            \node[anchor=north west, inner sep=0] (img) at (0,0) {\includegraphics[height=2.81cm]{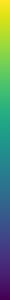}};
            \node[right] at (.1,0) {0.20};
            \node[right] at (.1,-2.81) {0};
        \end{tikzpicture}
      }
    \end{subfigure}

    \end{tabular}
   \caption{Comparisons of rendering using the Sobol sequence and using our SZ sequence with the \emph{Hangar Interior} environment map.
      Rendering was at 64spp with direct lighting only.
      Crops show representative results; full images are available in our supplemental material.
    }
    \label{fig:sportscar-hangar-interior}
\end{figure*}


Beyond these baseline tests, we implemented a more detailed test in which we evaluated Sobol and SZ points for rendering direct illumination from an environment map, using two error metrics: MRSE and the perceptually-based \FLIP metric~\cite{Andersson2020FLIP,Andersson2021FLIP}.
For this evaluation, however, we modified \emph{pbrt} so that the first two dimensions of the sequence are used to select a point inside each pixel area,
the next two are used to select a point on the lens,
the next two to importance-sample a direction from the environment map's distribution,
and then two more are used to sample the BSDF.
Note that this differs from \emph{pbrt}'s default assignment of dimensions for integration which does not
consider the possibility of $(0,2^q)$-sequences and so consumes additional 1D samples such as for time and sampling
which light source to sample from early dimensions of samplers.
We chose the \emph{Sportscar} scene for the evaluation since it includes a variety of complex measured BSDFs, ranging from nearly diffuse to highly specular.

Fig.~\ref{fig:sportscar-hangar-interior} shows results with a high-dynamic range environment map, including crops of representative parts of the image rendered at 64spp.
The SZ sampler provides a meaningful reduction of $1.93\times$ for MRSE; this error reduction is visually evident in images.
In this case, we can see that SZ does not suffer from the characteristic checkerboard pattern of unconverged Sobol sampling.
See Section~\ref{sec:further-results} for additional results with this scene, including a variety of different environment maps, visualizations of error across entire images,
and error measurements at additional sampling rates.

\subsection{Discrepancy}

\begin{figure}
    \setlength{\unit}{(\columnwidth - \gap)/2}                        
    {\scriptsize \centering
    \begin{tabular*}{1\columnwidth}{@{}c@{\extracolsep{\fill}}c@{}}
        \includegraphics[width=1\unit]{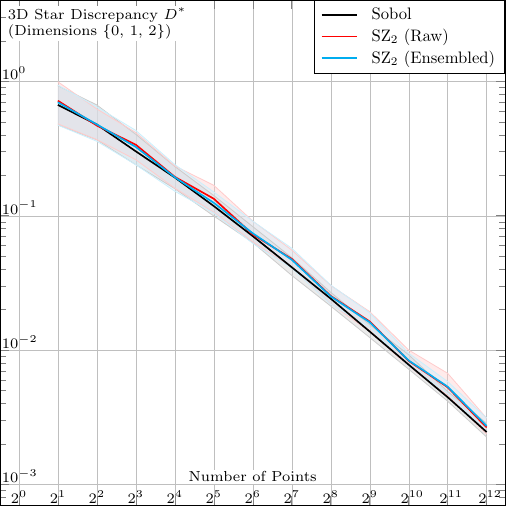}&%
        \includegraphics[width=1\unit]{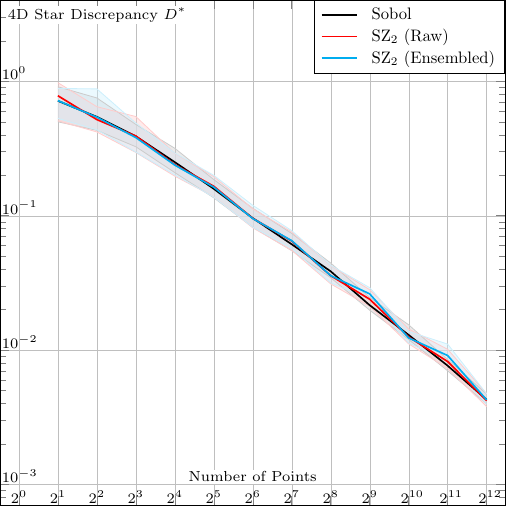}\\%
        (a) & (b) 
    \end{tabular*}
    }
    \caption{
        Star discrepancy in (a) three and (b) four dimensions, comparing raw and ensembled SZ sequences to Sobol, measured over 100 Owen-scrambled realizations.
        All the three sequences swing about close averages, with Sobol slightly better, but SZ seems to slightly  excel at powers of 4 in 4D.
    }
    \label{fig:discrepancy}
\end{figure}

The essence of Niederreiter's framework~\shortcite[Chapter~4]{Niederreiter92Random} is that identifying a sequence like SZ as $(0, 2^q)$-sequences in base $2^q$ is sufficient to qualify it as a low discrepancy sequence. We nontheless ran a test in the feasible range to empirically validate our model, and obtained the results in Fig.~\ref{fig:discrepancy}, confirming the theoretical model.
Please note that we skipped the first two dimensions, as they are identical to Sobol, Faure, and Niederreter.

\subsection{Frequency Spectra}

In Fig.~\ref{fig:teaser} we see a comparison between frequency spectra of corresponding pairs of dimensions between SZ and Sobol sequences.
Our construction evidently admits better pairwise projections at powers of $2^q$ octaves, though the discrepancy plots in Fig.~\ref{fig:discrepancy} suggest that Sobol behaves better at a more granular powers-of-2 octaves.
Further, the fact that all consecutive pairs of dimensions of ensembled SZ sequences are dyadic $(0, 2)$-sequences makes them ripe for different optimizations of such sequences like those suggested by Ahmed et al.~\shortcite{Ahmed23DigitalSequences, Ahmed2024SobolFast} or Doignies et. al. \shortcite{Doignies2025Differentiable}.
In the supplementary materials we provide more detailed spectral plots, as well as a tool to produce more.

\subsection{Time and Space Complexity}

As a binary-matrix-based construction, the baseline for time and space complexity is that of Sobol sequences, and migrating from Sobol in existing systems is as simple as replacing the matrices.
Notably, the matrices are identical for the the first two dimensions, hence preserving any special code for inversions \cite{Gruenschlos12Enumerating} or fast execution \cite{Ahmed2024SobolFast}.

\begin{figure}
    \setlength{\unit}{0.54\columnwidth}                        
    {\scriptsize \centering
    \begin{tabular*}{1\columnwidth}{@{}c@{\extracolsep{\fill}}c@{}}
        \includegraphics[height=1\unit]{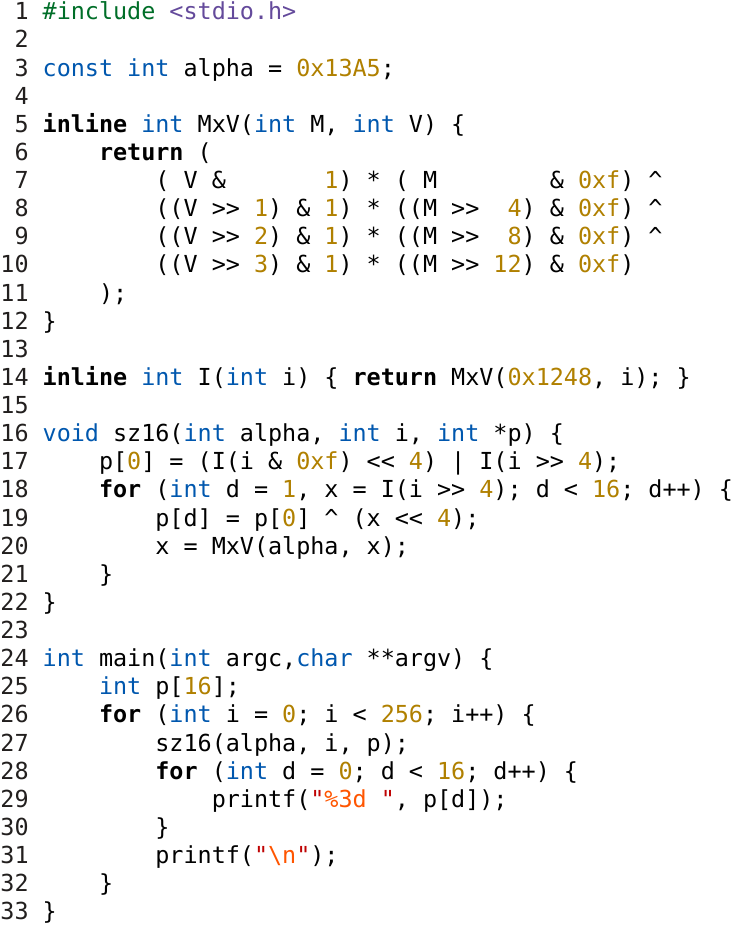}&%
        \includegraphics[height=1\unit]{images/compact/plot.pdf}\\%
        (a) Code & (b) plots of generated points
    \end{tabular*}
    }
    \caption{
        (a) Self-contained code example to compute 256 16D points that are pair-wise stratified in all 2D projections.
        (b) Plots to visualize the generated points without{\textbackslash}with Owen scrambling.
        The diagonal shows the implicit matrices used to compute the points in the dimension of respective rows and columns.
    }
    \label{fig:compact}
\end{figure}

Understanding the anatomy of SZ matrices, however, offers multiple ways to optimize the space and/or time complexity.
For a non-trivial example, in the code listing of Fig.~\ref{fig:compact}(a), we exploit the alpha property to generate a complete set of 256 pair-wise stratified points in 16D, as visualized in Fig.~\ref{fig:compact}(b).
In this example, the space complexity is reduced to a single integer \texttt{0x13A5} representing the alpha symbol of the alphabet in Fig.~\ref{fig:teaser}(b),
while other symbols are evaluated implicitly by computing subsequent dimensions from preceding ones.
Combined with Owen/Xor scrambling, this could represent a viable GPU-friendly sampling solutions for some applications, or a self-contained sample generator for hardware integration and/or in embedded systems.
The mentioned example is limited to two octaves and without nesting, but similar specialized solution might be tailored for more advanced requirements.

Apart from such specialized hacks and tricks, SZ sequences admit a generic optimization via this ``S-P-Z decomposition'':
\begin{equation}
    \mathcal{P}(a) = 
    \begin{pmatrix}
        I   &        &                 \\
            & a^{-1} &        &        \\
            &        & a^{-2} &        \\
            &        &        & \ddots \\
    \end{pmatrix}
    \mathcal{P}(I)
    \begin{pmatrix}
        I   &        &                 \\
            & a      &        &        \\
            &        & a^{2}  &        \\
            &        &        & \ddots \\
    \end{pmatrix} \,. \label{eq:spz-decomposition}
\end{equation}
The middle Pascal factor may be evaluated using the fast diagonal factoring introduced by Ahmed~\shortcite{Ahmed2024SobolFast}, while the block-diagonal factors are simply treated as a sum of diagonal matrices that translate similarly to bit shift-n-mask operations.
This reduces the time and space complexity from $\mathcal{O}(m)$ for $m$-bit precision to $\mathcal{O}(\log(m))$ for the middle $P$ factor and $\mathcal{O}(2q - 1)$ for the $S$ and $Z$ factors, independent of numeric precision, which is especially significant for 64-bit computations.
We actually obtained 11/23\% memory saving and 20/15$\times$ speedup for the first 16/256 dimensions, respectively.
An implementation is available in our supplementary materials. 

\subsection{Coding Complexity}

Even though ending up in only a few lines of code, implementing low-discrepancy construction to compute generator matrices is relatively difficult and quite prone to mistakes, and SZ sequences are no exception.
Acknowledging this, we will make our code publicly available, and aim at maintaining libraries for C, Python, etc., to encourage the adoption of these sequences.
That said, we still encourage the readers to try to implement the sequences themselves, as it can potentially improve understanding of their properties.

\section{Conclusion}

In this paper, we enriched the sampling library with a novel construction of binary-based low-discrepancy sequences that scale flexibly with dimension, offering multiple degrees of freedom and control to enable sampling solution architects to build sequences tailored to their needs.
Started experimentally, our empirical-based modeling led us close to the very well-established constructions of Faure and Niederreiter, with the primary difference that we replace elementary fields over (power-of-) prime bases by synthetic fields built over binary matrices, which enables very efficient computation.
More importantly, our model uniquely offers the capability of ensembling a high-dimensional LD sequence from low-dimensional sub-sequences, which is desirable in graphics applications like rendering, where the integration domain is mostly made of 2D surfaces.

Our sequences are evidently superior to Halton's for CG applications, and through multiple abstract and rendering tests, we demonstrated that they benchmark competitively with respect to the well-established Sobol sequences, offering a promising alternative to break their monopoly.
Sobol seems to lead in some measures like discrepancy, while SZ excels in others like spectral control and speed performance.
We believe, however, that a deeper analysis of the integrator-sampler interaction is needed before arriving at a final conclusion about the performance of these sequences.
We therefore expect this work to spark future research aiming at analyzing, improving, and utilizing these sequences.


\bibliographystyle{ACM-Reference-Format}
\bibliography{SZ.bib}


\begin{thebibliography}{40}


\ifx \showCODEN    \undefined \def \showCODEN     #1{\unskip}     \fi
\ifx \showDOI      \undefined \def \showDOI       #1{#1}\fi
\ifx \showISBNx    \undefined \def \showISBNx     #1{\unskip}     \fi
\ifx \showISBNxiii \undefined \def \showISBNxiii  #1{\unskip}     \fi
\ifx \showISSN     \undefined \def \showISSN      #1{\unskip}     \fi
\ifx \showLCCN     \undefined \def \showLCCN      #1{\unskip}     \fi
\ifx \shownote     \undefined \def \shownote      #1{#1}          \fi
\ifx \showarticletitle \undefined \def \showarticletitle #1{#1}   \fi
\ifx \showURL      \undefined \def \showURL       {\relax}        \fi
\providecommand\bibfield[2]{#2}
\providecommand\bibinfo[2]{#2}
\providecommand\natexlab[1]{#1}
\providecommand\showeprint[2][]{arXiv:#2}

\bibitem[Ahmed(2024)]%
        {Ahmed2024SobolFast}
\bibfield{author}{\bibinfo{person}{Abdalla G.~M. Ahmed}.}
  \bibinfo{year}{2024}\natexlab{}.
\newblock \showarticletitle{{An Implementation Algorithm of 2D Sobol Sequence
  Fast, Elegant, and Compact}}. In \bibinfo{booktitle}{\emph{Eurographics
  Symposium on Rendering}}, \bibfield{editor}{\bibinfo{person}{Eric Haines}
  {and} \bibinfo{person}{Elena Garces}} (Eds.). \bibinfo{publisher}{The
  Eurographics Association}.
\newblock
\showISBNx{978-3-03868-262-2}
\showISSN{1727-3463}
\urldef\tempurl%
\url{https://doi.org/10.2312/sr.20241147}
\showDOI{\tempurl}


\bibitem[Ahmed et~al\mbox{.}(2016)]%
        {Ahmed16LDBN}
\bibfield{author}{\bibinfo{person}{Abdalla G.~M. Ahmed},
  \bibinfo{person}{H{\'e}l\`{e}ne Perrier}, \bibinfo{person}{David Coeurjolly},
  \bibinfo{person}{Victor Ostromoukhov}, \bibinfo{person}{Jianwei Guo},
  \bibinfo{person}{Dong-Ming Yan}, \bibinfo{person}{Hui Huang}, {and}
  \bibinfo{person}{Oliver Deussen}.} \bibinfo{year}{2016}\natexlab{}.
\newblock \showarticletitle{{Low-Discrepancy Blue-Noise Sampling}}.
\newblock \bibinfo{journal}{\emph{ACM Trans. Graph.}} \bibinfo{volume}{35},
  \bibinfo{number}{6}, Article \bibinfo{articleno}{247} (\bibinfo{date}{Nov.}
  \bibinfo{year}{2016}), \bibinfo{numpages}{13}~pages.
\newblock
\showISSN{0730-0301}
\urldef\tempurl%
\url{https://doi.org/10.1145/2980179.2980218}
\showDOI{\tempurl}


\bibitem[Ahmed et~al\mbox{.}(2023)]%
        {Ahmed23DigitalSequences}
\bibfield{author}{\bibinfo{person}{Abdalla G.~M. Ahmed},
  \bibinfo{person}{Mikhail Skopenkov}, \bibinfo{person}{Markus Hadwiger}, {and}
  \bibinfo{person}{Peter Wonka}.} \bibinfo{year}{2023}\natexlab{}.
\newblock \showarticletitle{{Analysis and Synthesis of Digital Dyadic
  Sequences}}.
\newblock \bibinfo{journal}{\emph{ACM Trans. Graph.}} \bibinfo{volume}{42},
  \bibinfo{number}{6}, Article \bibinfo{articleno}{218} (\bibinfo{date}{Dec.}
  \bibinfo{year}{2023}), \bibinfo{numpages}{17}~pages.
\newblock
\urldef\tempurl%
\url{https://doi.org/10.1145/3618308}
\showDOI{\tempurl}


\bibitem[Ahmed and Wonka(2020)]%
        {Ahmed2020Screen}
\bibfield{author}{\bibinfo{person}{Abdalla G.~M. Ahmed} {and}
  \bibinfo{person}{Peter Wonka}.} \bibinfo{year}{2020}\natexlab{}.
\newblock \showarticletitle{Screen-space blue-noise diffusion of monte carlo
  sampling error via hierarchical ordering of pixels}.
\newblock \bibinfo{journal}{\emph{ACM Trans. Graph.}} \bibinfo{volume}{39},
  \bibinfo{number}{6}, Article \bibinfo{articleno}{244} (\bibinfo{date}{Nov.}
  \bibinfo{year}{2020}), \bibinfo{numpages}{15}~pages.
\newblock
\showISSN{0730-0301}
\urldef\tempurl%
\url{https://doi.org/10.1145/3414685.3417881}
\showDOI{\tempurl}


\bibitem[Ahmed and Wonka(2021)]%
        {Ahmed2021Optimizing}
\bibfield{author}{\bibinfo{person}{Abdalla G.~M. Ahmed} {and}
  \bibinfo{person}{Peter Wonka}.} \bibinfo{year}{2021}\natexlab{}.
\newblock \showarticletitle{{Optimizing Dyadic Nets}}.
\newblock \bibinfo{journal}{\emph{ACM Trans. Graph.}} \bibinfo{volume}{40},
  \bibinfo{number}{4}, Article \bibinfo{articleno}{141} (\bibinfo{date}{jul}
  \bibinfo{year}{2021}), \bibinfo{numpages}{17}~pages.
\newblock
\showISSN{0730-0301}
\urldef\tempurl%
\url{https://doi.org/10.1145/3450626.3459880}
\showDOI{\tempurl}


\bibitem[Andersson et~al\mbox{.}(2020)]%
        {Andersson2020FLIP}
\bibfield{author}{\bibinfo{person}{Pontus Andersson}, \bibinfo{person}{Jim
  Nilsson}, \bibinfo{person}{Tomas Akenine{-}M{\"{o}}ller},
  \bibinfo{person}{Magnus Oskarsson}, \bibinfo{person}{Kalle {\AA}str{\"{o}}m},
  {and} \bibinfo{person}{Mark~D. Fairchild}.} \bibinfo{year}{2020}\natexlab{}.
\newblock \showarticletitle{{{\FLIP:} {A} Difference Evaluator for Alternating
  Images}}.
\newblock \bibinfo{journal}{\emph{Proceedings of the ACM on Computer Graphics
  and Interactive Techniques}} \bibinfo{volume}{3}, \bibinfo{number}{2}
  (\bibinfo{year}{2020}), \bibinfo{pages}{15:1--15:23}.
\newblock
\urldef\tempurl%
\url{https://doi.org/10.1145/3406183}
\showDOI{\tempurl}


\bibitem[Andersson et~al\mbox{.}(2021)]%
        {Andersson2021FLIP}
\bibfield{author}{\bibinfo{person}{Pontus Andersson}, \bibinfo{person}{Jim
  Nilsson}, \bibinfo{person}{Peter Shirley}, {and} \bibinfo{person}{Tomas
  Akenine{-}M{\"{o}}ller}.} \bibinfo{year}{2021}\natexlab{}.
\newblock \showarticletitle{{Visualizing Errors in Rendered High Dynamic Range
  Images}}. In \bibinfo{booktitle}{\emph{Eurographics Short Papers}}.
\newblock
\urldef\tempurl%
\url{https://doi.org/10.2312/egs.20211015}
\showDOI{\tempurl}


\bibitem[Bratley et~al\mbox{.}(1992)]%
        {Bratley1992Implementation}
\bibfield{author}{\bibinfo{person}{Paul Bratley}, \bibinfo{person}{Bennett~L.
  Fox}, {and} \bibinfo{person}{Harald Niederreiter}.}
  \bibinfo{year}{1992}\natexlab{}.
\newblock \showarticletitle{{Implementation and Tests of Low-Discrepancy
  Sequences}}.
\newblock \bibinfo{journal}{\emph{ACM Trans. Model. Comput. Simul.}}
  \bibinfo{volume}{2}, \bibinfo{number}{3} (\bibinfo{date}{July}
  \bibinfo{year}{1992}), \bibinfo{pages}{195–213}.
\newblock
\showISSN{1049-3301}
\urldef\tempurl%
\url{https://doi.org/10.1145/146382.146385}
\showDOI{\tempurl}


\bibitem[Burley(2020)]%
        {Burley2020Scrambling}
\bibfield{author}{\bibinfo{person}{Brent Burley}.}
  \bibinfo{year}{2020}\natexlab{}.
\newblock \showarticletitle{{Practical Hash-based Owen Scrambling}}.
\newblock \bibinfo{journal}{\emph{Journal of Computer Graphics Techniques
  (JCGT)}} \bibinfo{volume}{10}, \bibinfo{number}{4} (\bibinfo{date}{29
  December} \bibinfo{year}{2020}), \bibinfo{pages}{1--20}.
\newblock
\showISSN{2331-7418}
\urldef\tempurl%
\url{http://jcgt.org/published/0009/04/01/}
\showURL{%
\tempurl}


\bibitem[Christensen et~al\mbox{.}(2018)]%
        {Christensen18Progressive}
\bibfield{author}{\bibinfo{person}{Per Christensen}, \bibinfo{person}{Andrew
  Kensler}, {and} \bibinfo{person}{Charlie Kilpatrick}.}
  \bibinfo{year}{2018}\natexlab{}.
\newblock \showarticletitle{{Progressive Multi-Jittered Sample Sequences}}. In
  \bibinfo{booktitle}{\emph{Computer Graphics Forum}},
  Vol.~\bibinfo{volume}{37}. Wiley Online Library, \bibinfo{pages}{21--33}.
\newblock


\bibitem[Cook et~al\mbox{.}(1984)]%
        {Cook1984Distributed}
\bibfield{author}{\bibinfo{person}{Robert~L. Cook}, \bibinfo{person}{Thomas
  Porter}, {and} \bibinfo{person}{Loren Carpenter}.}
  \bibinfo{year}{1984}\natexlab{}.
\newblock \showarticletitle{{Distributed Ray Tracing}}. In
  \bibinfo{booktitle}{\emph{Proceedings of the 11th Annual Conference on
  Computer Graphics and Interactive Techniques}}
  \emph{(\bibinfo{series}{SIGGRAPH '84})}. \bibinfo{publisher}{Association for
  Computing Machinery}, \bibinfo{address}{New York, NY, USA},
  \bibinfo{pages}{137–145}.
\newblock
\showISBNx{0897911385}
\urldef\tempurl%
\url{https://doi.org/10.1145/800031.808590}
\showDOI{\tempurl}


\bibitem[Doignies et~al\mbox{.}(2024)]%
        {Doignies2025Differentiable}
\bibfield{author}{\bibinfo{person}{Bastien Doignies}, \bibinfo{person}{David
  Coeurjolly}, \bibinfo{person}{Nicolas Bonneel}, \bibinfo{person}{Julie
  Digne}, \bibinfo{person}{Jean-Claude Iehl}, {and} \bibinfo{person}{Victor
  Ostromoukhov}.} \bibinfo{year}{2024}\natexlab{}.
\newblock \showarticletitle{{Differentiable Owen Scrambling}}.
\newblock \bibinfo{journal}{\emph{ACM Trans. Graph.}} \bibinfo{volume}{43},
  \bibinfo{number}{6}, Article \bibinfo{articleno}{255} (\bibinfo{date}{Nov.}
  \bibinfo{year}{2024}), \bibinfo{numpages}{12}~pages.
\newblock
\showISSN{0730-0301}
\urldef\tempurl%
\url{https://doi.org/10.1145/3687764}
\showDOI{\tempurl}


\bibitem[Faure(1982)]%
        {Faure1982Discrepancy}
\bibfield{author}{\bibinfo{person}{Henri Faure}.}
  \bibinfo{year}{1982}\natexlab{}.
\newblock \showarticletitle{Discrépance de Suites Associées à un Système de
  Numération (en Dimension s)}.
\newblock \bibinfo{journal}{\emph{Acta Arithmetica}} \bibinfo{volume}{41},
  \bibinfo{number}{4} (\bibinfo{year}{1982}), \bibinfo{pages}{337--351}.
\newblock
\urldef\tempurl%
\url{http://eudml.org/doc/205851}
\showURL{%
\tempurl}


\bibitem[Faure and Tezuka(2002)]%
        {Faure02Another}
\bibfield{author}{\bibinfo{person}{Henri Faure} {and} \bibinfo{person}{Shu
  Tezuka}.} \bibinfo{year}{2002}\natexlab{}.
\newblock \showarticletitle{Another Random Scrambling of Digital
  (t,s)-Sequences}. In \bibinfo{booktitle}{\emph{Monte Carlo and Quasi-Monte
  Carlo Methods 2000}}, \bibfield{editor}{\bibinfo{person}{Kai-Tai Fang},
  \bibinfo{person}{Harald Niederreiter}, {and} \bibinfo{person}{Fred~J.
  Hickernell}} (Eds.). \bibinfo{publisher}{Springer Berlin Heidelberg},
  \bibinfo{address}{Berlin, Heidelberg}, \bibinfo{pages}{242--256}.
\newblock
\showISBNx{978-3-642-56046-0}


\bibitem[Glassner(1994)]%
        {Glassner1995Principles}
\bibfield{author}{\bibinfo{person}{Andrew~S. Glassner}.}
  \bibinfo{year}{1994}\natexlab{}.
\newblock \bibinfo{booktitle}{\emph{Principles of Digital Image Synthesis}}.
\newblock \bibinfo{publisher}{Morgan Kaufmann Publishers Inc.},
  \bibinfo{address}{San Francisco, CA, USA}.
\newblock
\showISBNx{1558602763}


\bibitem[Gr{\"u}nschlo{\ss} et~al\mbox{.}(2012)]%
        {Gruenschlos12Enumerating}
\bibfield{author}{\bibinfo{person}{Leonhard Gr{\"u}nschlo{\ss}},
  \bibinfo{person}{Matthias Raab}, {and} \bibinfo{person}{Alexander Keller}.}
  \bibinfo{year}{2012}\natexlab{}.
\newblock \showarticletitle{{Enumerating Quasi-Monte Carlo Point Sequences in
  Elementary Intervals}}. In \bibinfo{booktitle}{\emph{Monte Carlo and
  Quasi-Monte Carlo Methods 2010}}, \bibfield{editor}{\bibinfo{person}{Leszek
  Plaskota} {and} \bibinfo{person}{Henryk Wo{\'{z}}niakowski}} (Eds.).
  \bibinfo{publisher}{Springer Berlin Heidelberg}, \bibinfo{address}{Berlin,
  Heidelberg}, \bibinfo{pages}{399--408}.
\newblock
\showISBNx{978-3-642-27440-4}


\bibitem[Halton(1960)]%
        {Halton1960efficiency}
\bibfield{author}{\bibinfo{person}{John~H Halton}.}
  \bibinfo{year}{1960}\natexlab{}.
\newblock \showarticletitle{On The Efficiency Of Certain Quasi-Random Sequences
  Of Points In Evaluating Multi-Dimensional Integrals}.
\newblock \bibinfo{journal}{\emph{Numer. Math.}}  \bibinfo{volume}{2}
  (\bibinfo{year}{1960}), \bibinfo{pages}{84--90}.
\newblock


\bibitem[Helmer et~al\mbox{.}(2021)]%
        {Helmer2021Stochastice}
\bibfield{author}{\bibinfo{person}{Andrew Helmer}, \bibinfo{person}{Per
  Christensen}, {and} \bibinfo{person}{Andrew Kensler}.}
  \bibinfo{year}{2021}\natexlab{}.
\newblock \showarticletitle{{Stochastic Generation of (t, s) Sample
  Sequences}}. In \bibinfo{booktitle}{\emph{Eurographics Symposium on Rendering
  - DL-only Track}}, \bibfield{editor}{\bibinfo{person}{Adrien Bousseau} {and}
  \bibinfo{person}{Morgan McGuire}} (Eds.). \bibinfo{publisher}{The
  Eurographics Association}.
\newblock
\showISBNx{978-3-03868-157-1}
\showISSN{1727-3463}
\urldef\tempurl%
\url{https://doi.org/10.2312/sr.20211287}
\showDOI{\tempurl}


\bibitem[Hošek and Wilkie(2012)]%
        {Hosek2012SkyDome}
\bibfield{author}{\bibinfo{person}{Lukáš Hošek} {and}
  \bibinfo{person}{Alexander Wilkie}.} \bibinfo{year}{2012}\natexlab{}.
\newblock \showarticletitle{{An Analytic Model for Full Spectral Sky-Dome
  Radiance}}.
\newblock \bibinfo{journal}{\emph{ACM Transactions on Graphics}}
  \bibinfo{volume}{31}, \bibinfo{number}{4}, Article \bibinfo{articleno}{95}
  (\bibinfo{date}{July} \bibinfo{year}{2012}), \bibinfo{numpages}{9}~pages.
\newblock
\showISSN{0730-0301}
\urldef\tempurl%
\url{https://doi.org/10.1145/2185520.2185591}
\showDOI{\tempurl}


\bibitem[Hošek and Wilkie(2013)]%
        {Hosek2013SolarRadiance}
\bibfield{author}{\bibinfo{person}{Lukáš Hošek} {and}
  \bibinfo{person}{Alexander Wilkie}.} \bibinfo{year}{2013}\natexlab{}.
\newblock \showarticletitle{{Adding a Solar-Radiance Function to the
  Hošek-Wilkie Skylight Model}}.
\newblock \bibinfo{journal}{\emph{IEEE Computer Graphics and Applications}}
  \bibinfo{volume}{33}, \bibinfo{number}{3} (\bibinfo{year}{2013}),
  \bibinfo{pages}{44--52}.
\newblock
\urldef\tempurl%
\url{https://doi.org/10.1109/MCG.2013.18}
\showDOI{\tempurl}


\bibitem[Inc.(2015)]%
        {OEISA258745}
\bibfield{author}{\bibinfo{person}{OEIS~Foundation Inc.}}
  \bibinfo{year}{2015}\natexlab{}.
\newblock \bibinfo{title}{Sequence A258745: Order of the general affine group
  AGL(n,2)}.
\newblock \bibinfo{howpublished}{The Online Encyclopedia of Integer Sequences}.
\newblock
\urldef\tempurl%
\url{https://oeis.org/A258745}
\showURL{%
\tempurl}
\newblock
\shownote{Accessed: 2025-01-18}.


\bibitem[Jarosz et~al\mbox{.}(2019)]%
        {Jarosz19Orthogonal}
\bibfield{author}{\bibinfo{person}{Wojciech Jarosz}, \bibinfo{person}{Afnan
  Enayet}, \bibinfo{person}{Andrew Kensler}, \bibinfo{person}{Charlie
  Kilpatrick}, {and} \bibinfo{person}{Per Christensen}.}
  \bibinfo{year}{2019}\natexlab{}.
\newblock \showarticletitle{{Orthogonal Array Sampling for Monte Carlo
  Rendering}}. In \bibinfo{booktitle}{\emph{Computer Graphics Forum}},
  Vol.~\bibinfo{volume}{38}. Wiley Online Library, \bibinfo{pages}{135--147}.
\newblock


\bibitem[Joe and Kuo(2008)]%
        {Joe2008Constructing}
\bibfield{author}{\bibinfo{person}{Stephen Joe} {and}
  \bibinfo{person}{Frances~Y. Kuo}.} \bibinfo{year}{2008}\natexlab{}.
\newblock \showarticletitle{{Constructing Sobol Sequences with Better
  Two-Dimensional Projections}}.
\newblock \bibinfo{journal}{\emph{SIAM J. Sci. Comput.}} \bibinfo{volume}{30},
  \bibinfo{number}{5} (\bibinfo{date}{Aug.} \bibinfo{year}{2008}),
  \bibinfo{pages}{2635–2654}.
\newblock
\showISSN{1064-8275}
\urldef\tempurl%
\url{https://doi.org/10.1137/070709359}
\showDOI{\tempurl}


\bibitem[Kollig and Keller(2002)]%
        {Kollig02Efficient}
\bibfield{author}{\bibinfo{person}{Thomas Kollig} {and}
  \bibinfo{person}{Alexander Keller}.} \bibinfo{year}{2002}\natexlab{}.
\newblock \showarticletitle{{Efficient Multidimensional Sampling}}. In
  \bibinfo{booktitle}{\emph{Computer Graphics Forum}},
  Vol.~\bibinfo{volume}{21}. \bibinfo{pages}{557--563}.
\newblock


\bibitem[Kuipers and Niederreiter(1974)]%
        {Kuipers74Uniform}
\bibfield{author}{\bibinfo{person}{Lauwerens Kuipers} {and}
  \bibinfo{person}{Harald Niederreiter}.} \bibinfo{year}{1974}\natexlab{}.
\newblock \bibinfo{booktitle}{\emph{{Uniform Distribution of Sequences}}}.
\newblock \bibinfo{publisher}{John Wiley \& Sons}, \bibinfo{address}{New York}.
\newblock
\showISBNx{0-471-51045-9}
\urldef\tempurl%
\url{http://opac.inria.fr/record=b1083239}
\showURL{%
\tempurl}
\newblock
\shownote{A Wiley-Interscience publication.}.


\bibitem[Mitchell(1992)]%
        {Mitchell1992ray}
\bibfield{author}{\bibinfo{person}{Don Mitchell}.}
  \bibinfo{year}{1992}\natexlab{}.
\newblock \showarticletitle{{Ray Tracing and Irregularities of Distribution}}.
  In \bibinfo{booktitle}{\emph{Third Eurographics Workshop on Rendering
  Proceedings}}. \bibinfo{pages}{61--69}.
\newblock


\bibitem[Niederreiter(1987)]%
        {Niederreiter87Point}
\bibfield{author}{\bibinfo{person}{Harald Niederreiter}.}
  \bibinfo{year}{1987}\natexlab{}.
\newblock \showarticletitle{Point Sets and Sequences with Small Discrepancy}.
\newblock \bibinfo{journal}{\emph{Monatshefte f{\"u}r Mathematik}}
  \bibinfo{volume}{104}, \bibinfo{number}{4} (\bibinfo{year}{1987}),
  \bibinfo{pages}{273--337}.
\newblock


\bibitem[Niederreiter(1992)]%
        {Niederreiter92Random}
\bibfield{author}{\bibinfo{person}{Harald Niederreiter}.}
  \bibinfo{year}{1992}\natexlab{}.
\newblock \bibinfo{booktitle}{\emph{{Random Number Generation and Quasi-Monte
  Carlo Methods}}}.
\newblock \bibinfo{publisher}{SIAM}.
\newblock


\bibitem[Ostromoukhov et~al\mbox{.}(2024)]%
        {Ostromoukhov2024QuadOptimized}
\bibfield{author}{\bibinfo{person}{Victor Ostromoukhov},
  \bibinfo{person}{Nicolas Bonneel}, \bibinfo{person}{David Coeurjolly}, {and}
  \bibinfo{person}{Jean-Claude Iehl}.} \bibinfo{year}{2024}\natexlab{}.
\newblock \showarticletitle{{Quad-Optimized Low-Discrepancy Sequences}}. In
  \bibinfo{booktitle}{\emph{ACM SIGGRAPH 2024 Conference Papers}} (Denver, CO,
  USA) \emph{(\bibinfo{series}{SIGGRAPH '24})}. \bibinfo{publisher}{Association
  for Computing Machinery}, \bibinfo{address}{New York, NY, USA}, Article
  \bibinfo{articleno}{72}, \bibinfo{numpages}{9}~pages.
\newblock
\showISBNx{9798400705250}
\urldef\tempurl%
\url{https://doi.org/10.1145/3641519.3657431}
\showDOI{\tempurl}


\bibitem[Owen(1995)]%
        {Owen95Randomly}
\bibfield{author}{\bibinfo{person}{Art~B. Owen}.}
  \bibinfo{year}{1995}\natexlab{}.
\newblock \showarticletitle{{Randomly Permuted (t,m,s)-Nets and (t,
  s)-Sequences}}. In \bibinfo{booktitle}{\emph{Monte Carlo and Quasi-Monte
  Carlo Methods in Scientific Computing}},
  \bibfield{editor}{\bibinfo{person}{Harald Niederreiter} {and}
  \bibinfo{person}{Peter Jau-Shyong Shiue}} (Eds.).
  \bibinfo{publisher}{Springer New York}, \bibinfo{address}{New York, NY},
  \bibinfo{pages}{299--317}.
\newblock
\showISBNx{978-1-4612-2552-2}


\bibitem[Paulin et~al\mbox{.}(2021)]%
        {Paulin2021Cascaded}
\bibfield{author}{\bibinfo{person}{Lo\"{\i}s Paulin}, \bibinfo{person}{David
  Coeurjolly}, \bibinfo{person}{Jean-Claude Iehl}, \bibinfo{person}{Nicolas
  Bonneel}, \bibinfo{person}{Alexander Keller}, {and} \bibinfo{person}{Victor
  Ostromoukhov}.} \bibinfo{year}{2021}\natexlab{}.
\newblock \showarticletitle{Cascaded Sobol' Sampling}.
\newblock \bibinfo{journal}{\emph{ACM Trans. Graph.}} \bibinfo{volume}{40},
  \bibinfo{number}{6}, Article \bibinfo{articleno}{275} (\bibinfo{date}{dec}
  \bibinfo{year}{2021}), \bibinfo{numpages}{13}~pages.
\newblock
\showISSN{0730-0301}
\urldef\tempurl%
\url{https://doi.org/10.1145/3478513.3480482}
\showDOI{\tempurl}


\bibitem[Perrier et~al\mbox{.}(2018)]%
        {Perrier18Sequences}
\bibfield{author}{\bibinfo{person}{H\'el\`ene Perrier}, \bibinfo{person}{David
  Coeurjolly}, \bibinfo{person}{Feng Xie}, \bibinfo{person}{Matt Pharr},
  \bibinfo{person}{Pat Hanrahan}, {and} \bibinfo{person}{Victor Ostromoukhov}.}
  \bibinfo{year}{2018}\natexlab{}.
\newblock \showarticletitle{{Sequences with Low-Discrepancy Blue-Noise 2-D
  Projections}}.
\newblock \bibinfo{journal}{\emph{Computer Graphics Forum (Proceedings of
  Eurographics)}} \bibinfo{volume}{37}, \bibinfo{number}{2}
  (\bibinfo{year}{2018}), \bibinfo{pages}{339–353}.
\newblock


\bibitem[Pharr(2019)]%
        {Pharr19Efficient}
\bibfield{author}{\bibinfo{person}{Matt Pharr}.}
  \bibinfo{year}{2019}\natexlab{}.
\newblock \showarticletitle{{Efficient Generation of Points that Satisfy
  Two-Dimensional Elementary Intervals}}.
\newblock \bibinfo{journal}{\emph{Journal of Computer Graphics Techniques
  (JCGT)}} \bibinfo{volume}{8}, \bibinfo{number}{1} (\bibinfo{date}{27
  February} \bibinfo{year}{2019}), \bibinfo{pages}{56--68}.
\newblock
\showISSN{2331-7418}
\urldef\tempurl%
\url{http://jcgt.org/published/0008/01/04/}
\showURL{%
\tempurl}


\bibitem[Pharr and Humphreys(2004)]%
        {Pharr04PBRT}
\bibfield{author}{\bibinfo{person}{Matt Pharr} {and} \bibinfo{person}{Greg
  Humphreys}.} \bibinfo{year}{2004}\natexlab{}.
\newblock \bibinfo{booktitle}{\emph{{Physically-Based Rendering: from Theory to
  Implementation}} (\bibinfo{edition}{1st} ed.)}.
\newblock \bibinfo{publisher}{Morgan Kaufmann Publishers Inc.}
\newblock
\showISBNx{9780123908568}


\bibitem[Pharr et~al\mbox{.}(2016)]%
        {Pharr16PBRT}
\bibfield{author}{\bibinfo{person}{Matt Pharr}, \bibinfo{person}{Wenzel Jakob},
  {and} \bibinfo{person}{Greg Humphreys}.} \bibinfo{year}{2016}\natexlab{}.
\newblock \bibinfo{booktitle}{\emph{{Physically Based Rendering: From Theory to
  Implementation}} (\bibinfo{edition}{3rd} ed.)}.
\newblock \bibinfo{publisher}{Morgan Kaufmann Publishers Inc.},
  \bibinfo{address}{San Francisco, CA, USA}.
\newblock
\showISBNx{0128006455}


\bibitem[Pharr et~al\mbox{.}(2023)]%
        {Pharr23PBRT}
\bibfield{author}{\bibinfo{person}{Matt Pharr}, \bibinfo{person}{Wenzel Jakob},
  {and} \bibinfo{person}{Greg Humphreys}.} \bibinfo{year}{2023}\natexlab{}.
\newblock \bibinfo{booktitle}{\emph{{Physically Based Rendering: From Theory to
  Implementation}} (\bibinfo{edition}{4th} ed.)}.
\newblock \bibinfo{publisher}{MIT Press}.
\newblock
\showISBNx{0262048027}


\bibitem[Sobol'(1967)]%
        {sobol1967}
\bibfield{author}{\bibinfo{person}{Il'ya~Meerovich Sobol'}.}
  \bibinfo{year}{1967}\natexlab{}.
\newblock \showarticletitle{{On the Distribution of Points in a Cube and the
  Approximate Evaluation of Integrals}}.
\newblock \bibinfo{journal}{\emph{Zhurnal Vychislitel'noi Matematiki i
  Matematicheskoi Fiziki}} \bibinfo{volume}{7}, \bibinfo{number}{4}
  (\bibinfo{year}{1967}), \bibinfo{pages}{784--802}.
\newblock


\bibitem[Tezuka(1994)]%
        {Tezuka1994Generalization}
\bibfield{author}{\bibinfo{person}{Shu Tezuka}.}
  \bibinfo{year}{1994}\natexlab{}.
\newblock \showarticletitle{A Generalization of Faure Sequences and its
  Efficient Implementation}.
\newblock \bibinfo{journal}{\emph{Technical Report, IBM Research, Tokyo
  Research Laboratory}} (\bibinfo{year}{1994}).
\newblock
\urldef\tempurl%
\url{https://doi.org/10.13140/RG.2.2.16748.16003}
\showDOI{\tempurl}


\bibitem[van~der Corput(1935)]%
        {vanderCorput1935}
\bibfield{author}{\bibinfo{person}{J.G. van~der Corput}.}
  \bibinfo{year}{1935}\natexlab{}.
\newblock \showarticletitle{{Verteilungsfunktionen}}.
\newblock \bibinfo{journal}{\emph{Proceedings of the Nederlandse Akademie van
  Wetenschappen}} \bibinfo{number}{38} (\bibinfo{year}{1935}),
  \bibinfo{pages}{813--821}.
\newblock


\bibitem[Warren(2012)]%
        {Warren2012Hacker}
\bibfield{author}{\bibinfo{person}{Henry~S. Warren}.}
  \bibinfo{year}{2012}\natexlab{}.
\newblock \bibinfo{booktitle}{\emph{{Hacker's Delight}} (\bibinfo{edition}{2nd}
  ed.)}.
\newblock \bibinfo{publisher}{Addison-Wesley Professional}.
\newblock
\showISBNx{0321842685}


\end{thebibliography}


\appendix

\section{Additional Results}
\label{sec:further-results}

\subsection{Numerical Integration}

\begin{figure}[b]
  \centering
    \begin{tabular}{c@{\hspace{0.1cm}}c}
    \begin{subfigure}[t]{0.44\linewidth}
      \centering
      \caption*{\hspace{0.7cm}$f^D_{1,2} * f^D_{3,4}$}
        \vbox{
            \includegraphics[width=\linewidth]{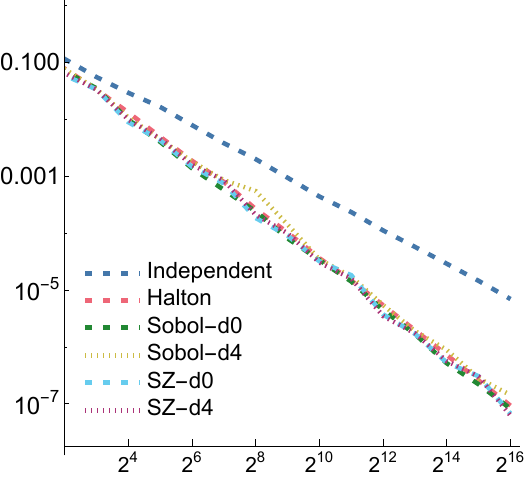}
        }
    \end{subfigure} &
    \begin{subfigure}[t]{0.44\linewidth}
      \centering
      \caption*{\hspace{0.5cm}$f^D_{1,2,3,4}$}
        \vbox{
            \includegraphics[width=\linewidth]{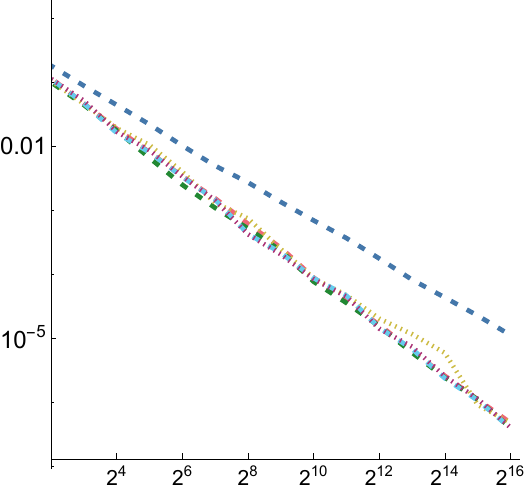}
        }
    \end{subfigure} \\
    \begin{subfigure}[t]{0.44\linewidth}
      \centering
      \caption*{\hspace{0.5cm}$f^D_{1,2} * f^D_{3,4} + f^D_{5,6} * f^D_{7,8}$}
        \vbox{
            \includegraphics[width=\linewidth]{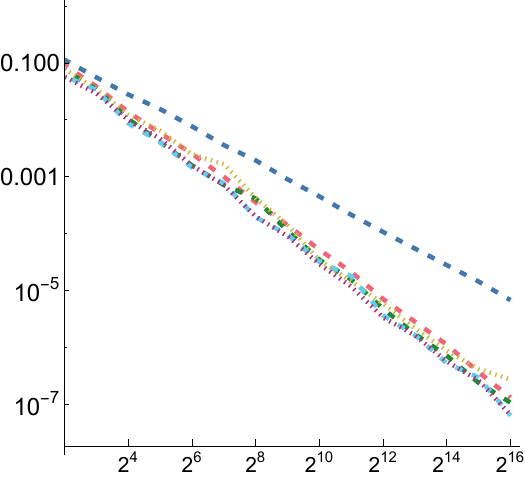}
        }
    \end{subfigure} &
   \begin{subfigure}[t]{0.44\linewidth}
      \centering
      \caption*{\hspace{0.7cm}$f^D_{1,2} * f^D_{1,3} * f^D_{1,4} * f^D_{2,3} * f^D_{2,4} * f^D_{3,4}$}
        \vbox{
            \includegraphics[width=\linewidth]{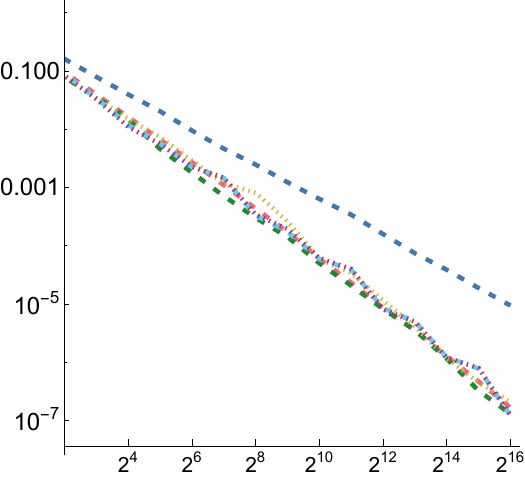}
        }
    \end{subfigure}

    \end{tabular}
    \caption{MRSE with $g^0$ with a variety of analytic 4- and 8-dimensional functions.
      As in Fig.~\ref{fig:analytic-function-error}, ``Sobol-d0'' and ``Sobol-d4'' denote Sobol points starting at dimension 0 and 4, respectively,
      and similarly for SZ.
    }
    \label{fig:g0-analytic-function-error}
\end{figure}

Fig.~\ref{fig:g0-analytic-function-error} presents MRSE measurements for four synthetic functions
using the $g^0$ function from Sec.~\ref{sec:integration}.
Note that unlike with the $g^1$ and $g^\infty$ functions, all considered LD sampling patterns have
similar error with the step function, though all still are substantially better than independent sampling.

\subsection{Rendering}

\begin{figure*}[tb]
  \centering
    \begin{tabular}{c@{\hspace{0.005cm}}c@{\hspace{0.075cm}}c@{\hspace{0.075cm}}c@{\hspace{0.075cm}}c@{\hspace{0.075cm}}c@{\hspace{0.075cm}}c}

    \raisebox{-0.95cm}{\rotatebox[origin=r]{90}{\textsc{Cayley Interior}}} &
    \begin{subfigure}[t]{0.3\linewidth}
      \centering
        \caption*{\textsc{Reference}}
        \vbox{
            \includegraphics[height=2.8cm]{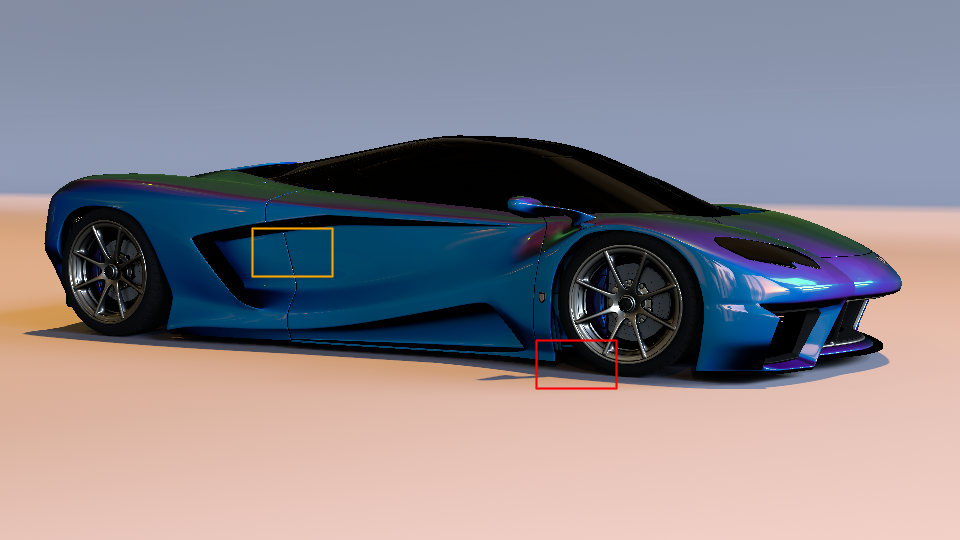}
        }
        \caption*{MRSE / \FLIP\\~}
    \end{subfigure} &
    \begin{subfigure}[t]{0.13\linewidth}
      \centering
         \caption*{\textsc{Sobol}}
         \vbox{
            \includegraphics[height=1.3cm]{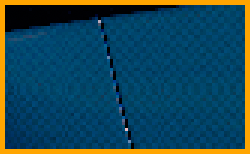}\vspace{0.2cm}
            \includegraphics[height=1.3cm]{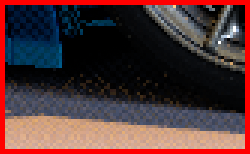}
        }
        \caption*{0.018208 / 0.037904\\~}
    \end{subfigure} &
    \begin{subfigure}[t]{0.13\linewidth}
      \centering
        \caption*{\textsc{Sobol MRSE}}
        \vbox{
            \includegraphics[height=1.3cm]{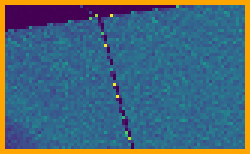}\vspace{0.2cm}
            \includegraphics[height=1.3cm]{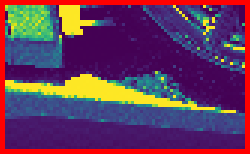}
        }
    \end{subfigure} &
    \begin{subfigure}[t]{0.13\linewidth}
      \centering
        \caption*{\textsc{SZ}}
        \vbox{
            \includegraphics[height=1.3cm]{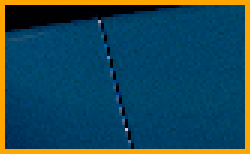}\vspace{0.2cm}
            \includegraphics[height=1.3cm]{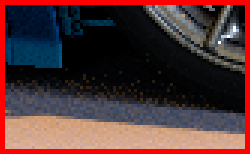}
        }
        \caption*{\textbf{0.012227} / \textbf{0.034762}\\~}
    \end{subfigure} &
    \begin{subfigure}[t]{0.13\linewidth}
      \centering
         \caption*{\textsc{SZ MRSE}}
       \vbox{
            \includegraphics[height=1.3cm]{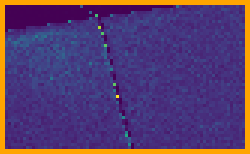}\vspace{0.2cm}
            \includegraphics[height=1.3cm]{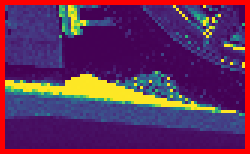}
        }
    \end{subfigure} &
    \begin{subfigure}[t]{0.13\linewidth}
        \centering
         \caption*{\textsc{Reference}}
       \vbox{
            \includegraphics[height=1.3cm]{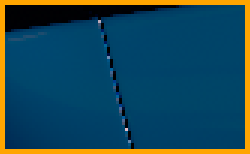}\vspace{0.2cm}
            \includegraphics[height=1.3cm]{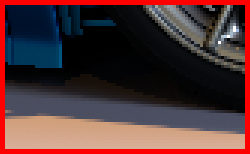}
        }
    \end{subfigure} \\

    \raisebox{2.55cm}{\rotatebox[origin=r]{90}{\textsc{Empty Warehouse}}} &
    \begin{subfigure}[t]{0.3\linewidth}
      \centering
        \vbox{
            \includegraphics[height=2.8cm]{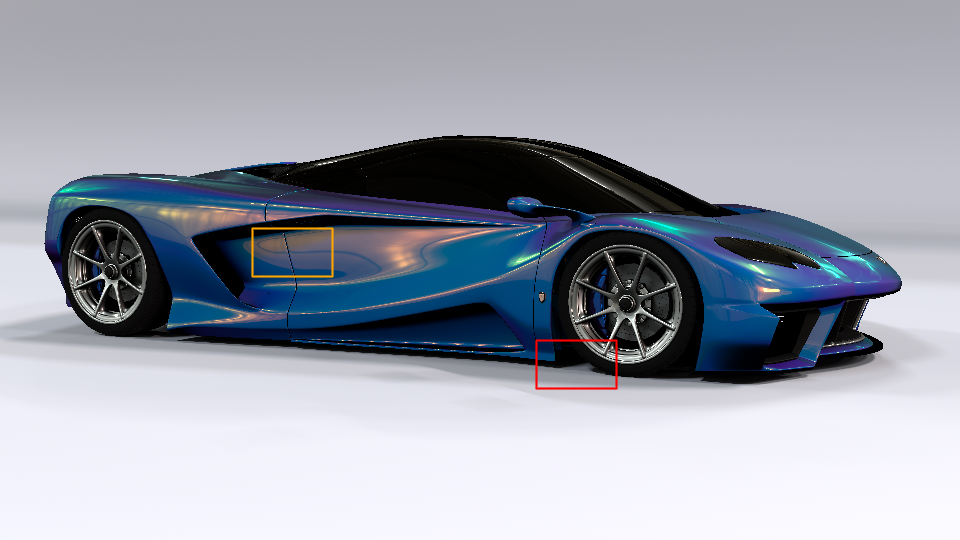}
        }
        \caption*{MRSE / \FLIP\\~}
    \end{subfigure} &
    \begin{subfigure}[t]{0.13\linewidth}
      \centering
        \vbox{
            \includegraphics[height=1.3cm]{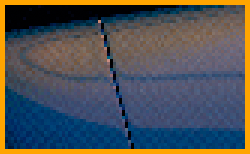}\vspace{0.2cm}
            \includegraphics[height=1.3cm]{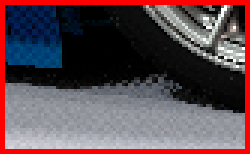}
        }
        \caption*{0.021288 / 0.035138\\~}
    \end{subfigure} &
    \begin{subfigure}[t]{0.13\linewidth}
      \centering
        \vbox{
            \includegraphics[height=1.3cm]{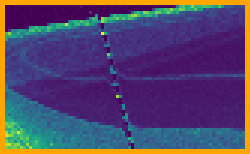}\vspace{0.2cm}
            \includegraphics[height=1.3cm]{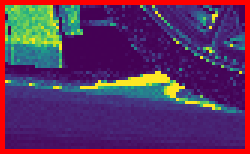}
        }
    \end{subfigure} &
    \begin{subfigure}[t]{0.13\linewidth}
      \centering
        \vbox{
            \includegraphics[height=1.3cm]{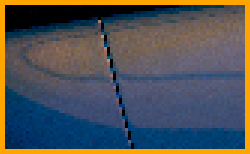}\vspace{0.2cm}
            \includegraphics[height=1.3cm]{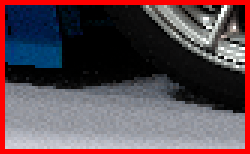}
        }
        \caption*{\textbf{0.013279} / \textbf{0.032293}\\~}
    \end{subfigure} &
    \begin{subfigure}[t]{0.13\linewidth}
      \centering
        \vbox{
            \includegraphics[height=1.3cm]{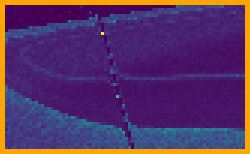}\vspace{0.2cm}
            \includegraphics[height=1.3cm]{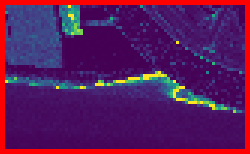}
        }
    \end{subfigure} &
    \begin{subfigure}[t]{0.13\linewidth}
        \centering
        \vbox{
            \includegraphics[height=1.3cm]{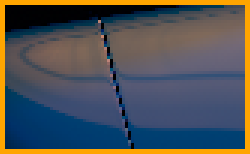}\vspace{0.2cm}
            \includegraphics[height=1.3cm]{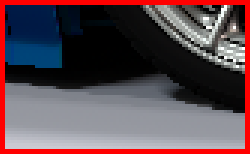}
        }
    \end{subfigure} \\

    \raisebox{1.5cm}{\rotatebox[origin=r]{90}{\textsc{Sky}}} &
    \begin{subfigure}[t]{0.3\linewidth}
      \centering
        \vbox{
            \includegraphics[height=2.8cm]{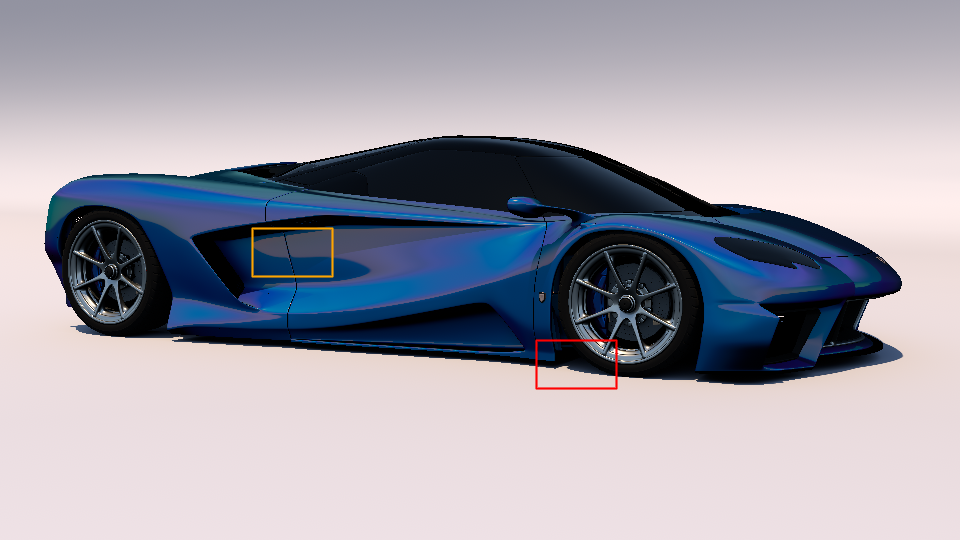}
        }
        \caption*{MRSE / \FLIP\\~}
    \end{subfigure} &
    \begin{subfigure}[t]{0.13\linewidth}
      \centering
        \vbox{
            \includegraphics[height=1.3cm]{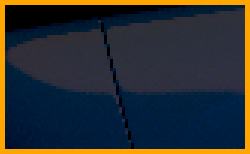}\vspace{0.2cm}
            \includegraphics[height=1.3cm]{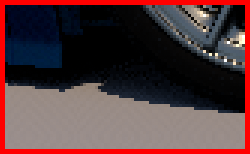}
        }
        \caption*{\textbf{0.001472} / \textbf{0.012582}\\~}
    \end{subfigure} &
    \begin{subfigure}[t]{0.13\linewidth}
      \centering
        \vbox{
            \includegraphics[height=1.3cm]{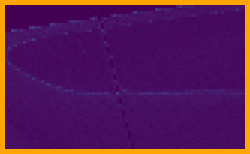}\vspace{0.2cm}
            \includegraphics[height=1.3cm]{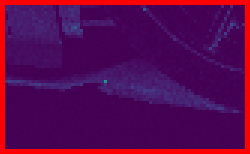}
        }
    \end{subfigure} &
    \begin{subfigure}[t]{0.13\linewidth}
      \centering
        \vbox{
            \includegraphics[height=1.3cm]{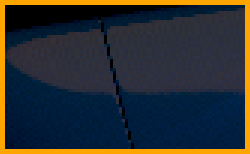}\vspace{0.2cm}
            \includegraphics[height=1.3cm]{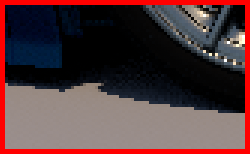}
        }
        \caption*{0.001641 / 0.012173\\~}
    \end{subfigure} &
    \begin{subfigure}[t]{0.13\linewidth}
      \centering
        \vbox{
            \includegraphics[height=1.3cm]{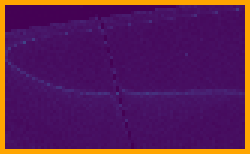}\vspace{0.2cm}
            \includegraphics[height=1.3cm]{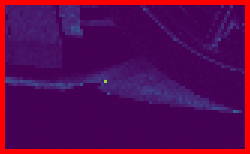}
        }
    \end{subfigure} &
    \begin{subfigure}[t]{0.13\linewidth}
        \centering
        \vbox{
            \includegraphics[height=1.3cm]{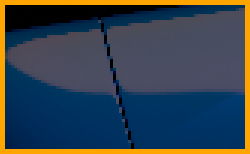}\vspace{0.2cm}
            \includegraphics[height=1.3cm]{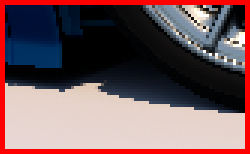}
        }
    \end{subfigure} \\

    \end{tabular}

    \caption{Comparisons of images rendered using the Sobol sequence and using our SZ sequence.
      All images are rendered with 64spp with direct lighting only, using four varied environment maps.
      Crops show representative results; full images are available in our supplemental material.
      See Figure~\ref{fig:mrseimages} for the specification of the heatmap used to visualize MRSE error.
    }
    \label{fig:sportscar-env-map-images}
\end{figure*}

Fig.~\ref{fig:sportscar-env-map-images} presents additional results for the \emph{Sportscar} scene with additional environment maps to supplement Fig.~\ref{fig:sportscar-hangar-interior}.
Together, we have the following environment maps, spanning a range of illumination:
\begin{itemize}
    \item \emph{Empty Warehouse}: the interior of a large warehouse, primarily lit by fluorescent lights but with a small window allowing some daylight.
    (Fig.~\ref{fig:sportscar-hangar-interior}).
    \item \emph{Cayley Interior}: the interior of a house, mostly illuminated with daylight through the doors of a balcony.
    \item \emph{Hangar Interior}: the interior of an airplane hangar, lit  by daylight both through skylights and a large open door.
    \item \emph{Sky}: an analytic sky model~\cite{Hosek2012SkyDome,Hosek2013SolarRadiance}.
\end{itemize}
For all of these (and for Fig.~\ref{fig:sportscar-hangar-interior}), we report the average error over 50 images rendered with different random seeds to average out small variations in error across independent runs.
Reference images were rendered with 32,768 samples per pixel (spp).

For the first two additional environment maps, SZ also reduces error compared to Sobol sampling;
\emph{Sky} has slightly higher error with SZ than with Sobol samples, though error for both is low.
In general, the greatest improvement seems to come with the most complex environment maps.
Given both a complex BSDF and complex illumination, sampling their full 4D product well is important for accurate rendering.
We hypothesize that the scenes with more complex environments see more improvement thanks to SZ providing a $(0,4)$-sequence for those dimensions.

\begin{figure*}[tb]
  \centering
  \begin{tabular}{c@{\hspace{0.2cm}}c@{\hspace{0.2cm}}c@{\hspace{0.2cm}}c@{\hspace{0.2cm}}c@{\hspace{0.2cm}}c}
    \raisebox{3.9cm}{\rotatebox[origin=r]{90}{\textsc{SZ}\hspace{1.9cm}\textsc{Sobol}}} &
    \begin{subfigure}[t]{0.22\linewidth}
        \centering
        \vbox{
          \includegraphics[height=2.25cm]{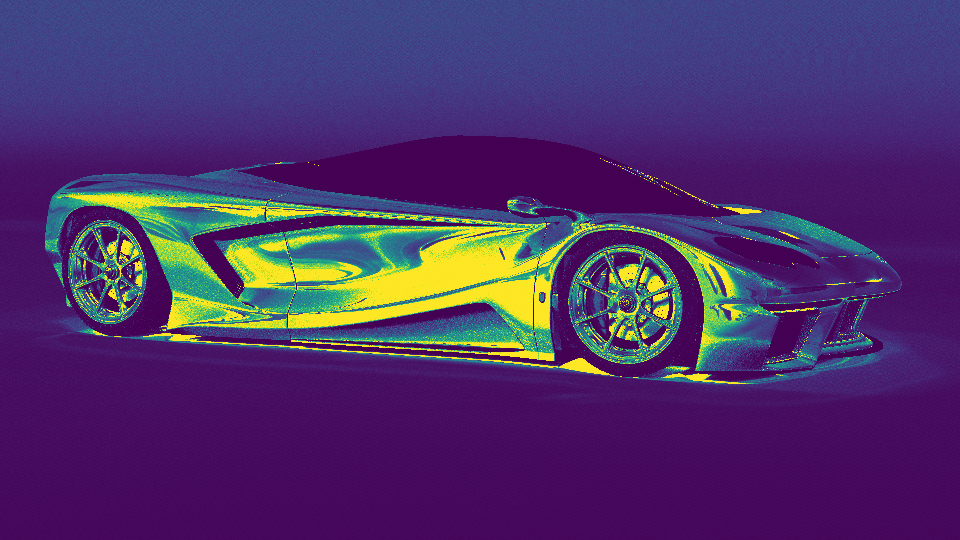}\vspace{0.1cm}
          \includegraphics[height=2.25cm]{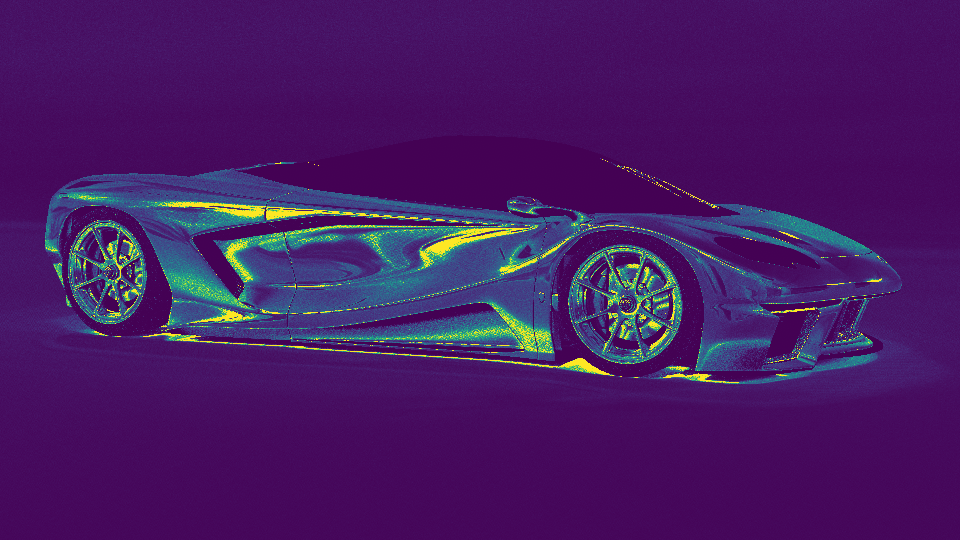}
        }
        \caption*{\textsc{Hangar Interior}}
    \end{subfigure} &
    \begin{subfigure}[t]{0.22\linewidth}
        \centering
        \vbox{
          \includegraphics[height=2.25cm]{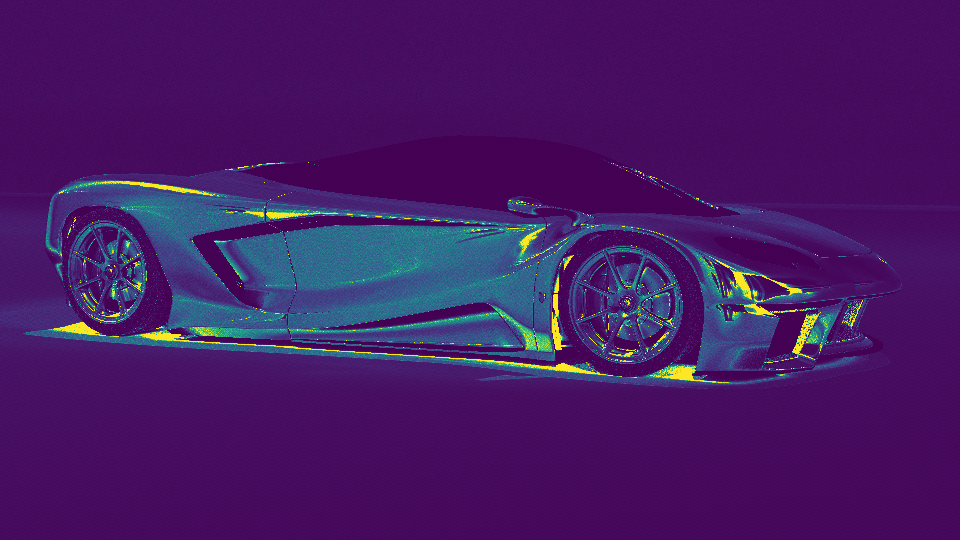}\vspace{0.1cm}
          \includegraphics[height=2.25cm]{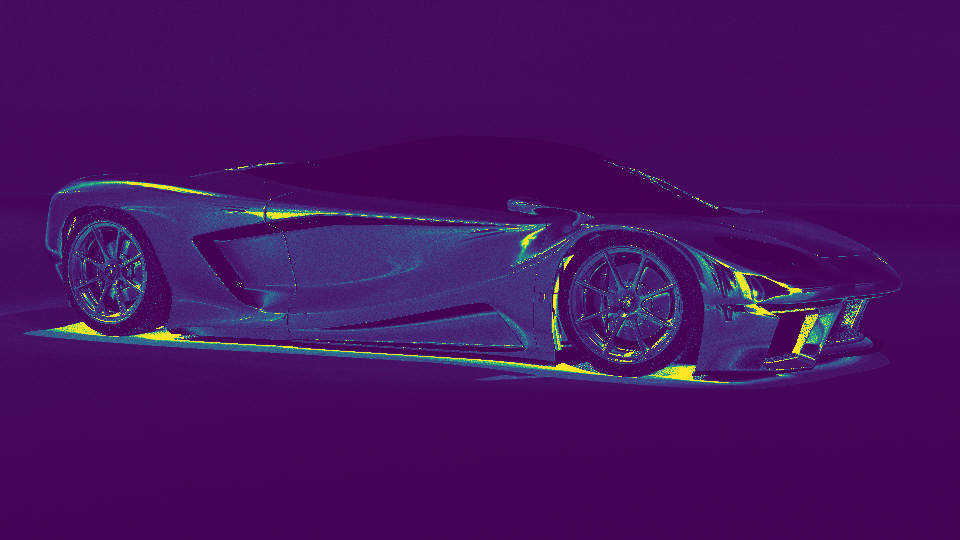}
        }
        \caption*{\textsc{Cayley Interior}}
    \end{subfigure} &
    \begin{subfigure}[t]{0.22\linewidth}
        \centering
        \vbox{
          \includegraphics[height=2.25cm]{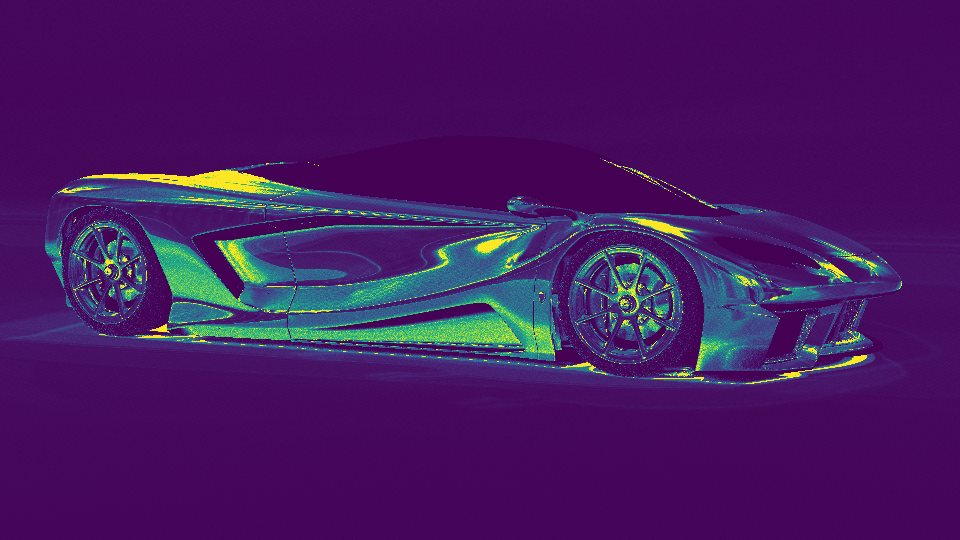}\vspace{0.1cm}
          \includegraphics[height=2.25cm]{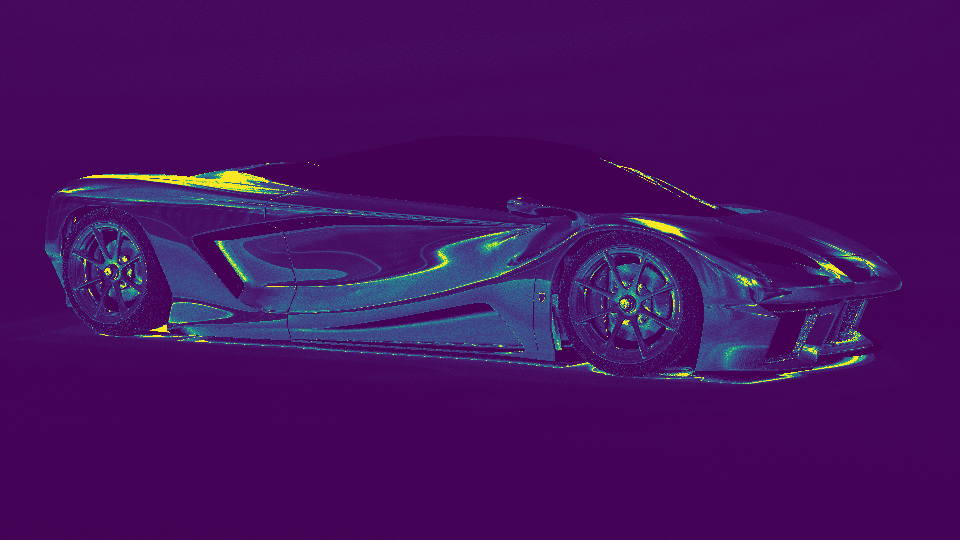}
        }
        \caption*{\textsc{Empty Warehouse}}
    \end{subfigure} &
    \begin{subfigure}[t]{0.22\linewidth}
        \centering
        \vbox{
          \includegraphics[height=2.25cm]{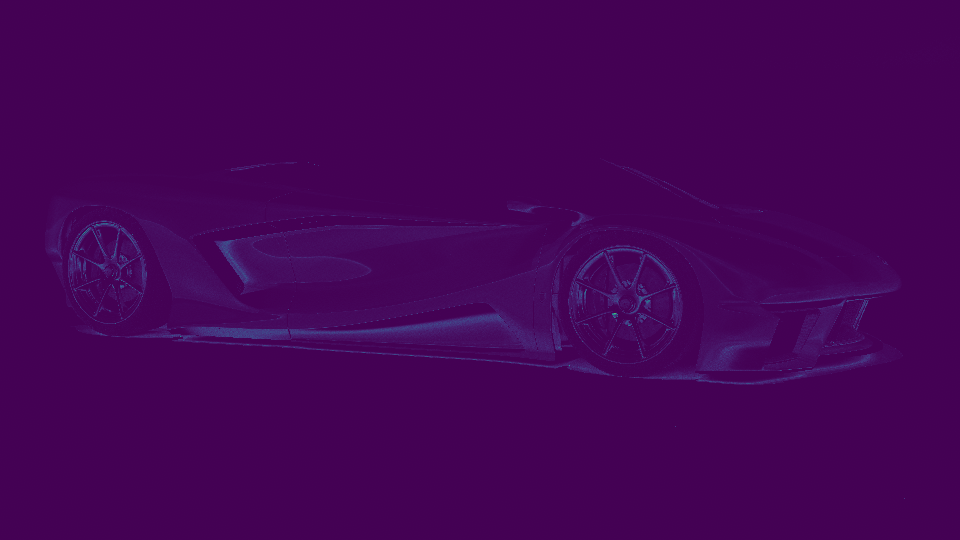}\vspace{0.1cm}
          \includegraphics[height=2.25cm]{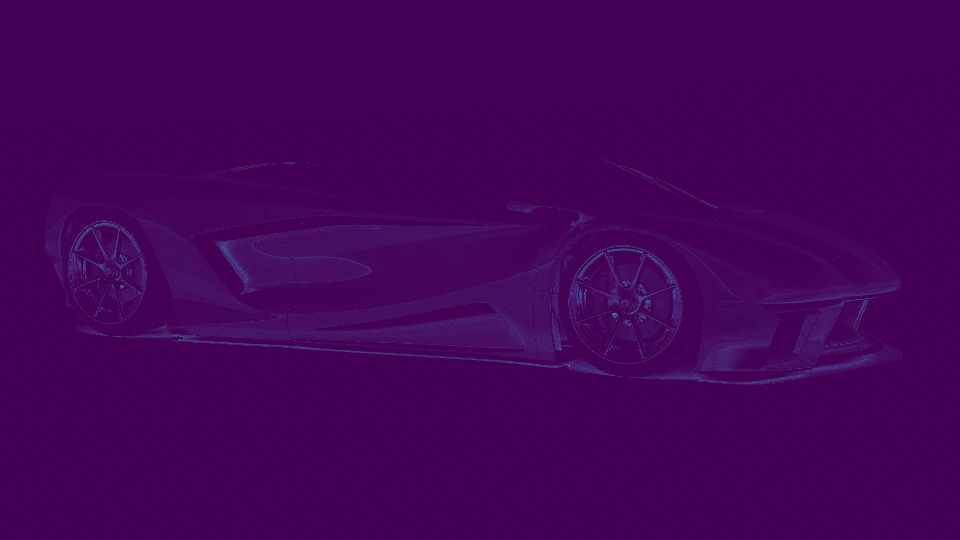}
        }
        \caption*{\textsc{Sky}}
    \end{subfigure} &
    \begin{subfigure}[t]{0.05\linewidth}
      \raisebox{0.2\height}{%
        \begin{tikzpicture}
            \node[anchor=north west, inner sep=0] (img) at (0,0) {\includegraphics[height=3cm]{images/heatmap.png}};
            \node[right] at (.1,0) {0.20};
            \node[right] at (.1,-3) {0};
        \end{tikzpicture}
      }
    \end{subfigure} \\
    
    \end{tabular}

    \caption{Visualizations of MRSE for the four scenes shown in Figure~\ref{fig:mrseimages},
      generated with test images rendered at 64spp.
      The benefit from SZ points is generally small in the relatively diffuse ground-plane and background; the largest reductions
      in error are on the car body, which has a complex BSDF.
    }
    \label{fig:mrseimages}
\end{figure*}

MRSE visualizations for the entire images with both samplers are shown in Figure~\ref{fig:mrseimages}.
We can see that the error reduction is greatest with the complex glossy specular BSDF of the car's paint, though
for some scenes there is also some benefit for the near-diffuse ground plane and background.

Error at different numbers of samples per pixel is reported in Tables~\ref{tab:rendered-mrse} and~\ref{tab:rendered-flip}.
SZ has lower MRSE than Sobol over a range of power-of-4 sampling rates and also has lower \FLIP error at low sampling rates.
At higher sampling rates, both have very low \FLIP error.
It is interesting to note that the relative performance of SZ and Sobol can be quite different at different sampling rates, even for the same scene.
This is somewhat unexpected, as we generally expect MRSE to decrease at a constant rate with increasing sampling rate.
We also see that with \emph{Sky}, SZ does extremely well at 16 and 256spp---both even powers of 4 samples.
At those rates, SZ has nearly $3\times$ lower error than Sobol sampling.

We have also measured performance of our sampler on a 32-core AMD 3970X CPU and an NVIDIA A6000 GPU.
We find that there is no meaningful performance difference when using the \texttt{SobolSampler} or our \texttt{SZSampler}; the time spent generating sample points is less than 3\% of total rendering time.

\begin{table*}[tb]
  \centering
  \begin{tabular}{r|rrr|rrr|rrr|rrr}
 & \multicolumn{3}{c|}{Hangar Interior} & \multicolumn{3}{c|}{Cayley Interior} & \multicolumn{3}{c|}{Empty Warehouse} & \multicolumn{3}{c}{Sky} \\
 spp & \multicolumn{1}{c}{Sobol} & \multicolumn{1}{c}{SZ} & \multicolumn{1}{c|}{Ratio} &
 \multicolumn{1}{c}{Sobol} & \multicolumn{1}{c}{SZ} & \multicolumn{1}{c|}{Ratio} &
 \multicolumn{1}{c}{Sobol} & \multicolumn{1}{c}{SZ} & \multicolumn{1}{c|}{Ratio} &
 \multicolumn{1}{c}{Sobol} & \multicolumn{1}{c}{SZ} & \multicolumn{1}{c}{Ratio} \\
\hline
16 &$0.128$ & $0.111$ & $1.15\times$ &
$0.0856$ & $0.0697$ & $1.23\times$ &
$0.111$ & $0.0803$ & $1.38\times$ &
$0.0329$ & $0.0111$ & $2.96\times$ \\
64 &$0.0372$ & $0.0193$ & $1.93\times$ &
$0.0182$ & $0.0122$ & $1.49\times$ &
$0.0213$ & $0.0133$ & $1.60\times$ &
$0.00147$ & $0.00164$ & $0.897\times$ \\
256 &$0.00383$ & $0.00393$ & $0.977\times$ &
$0.0029$ & $0.00216$ & $1.34\times$ &
$0.00297$ & $0.00227$ & $1.31\times$ &
$0.000412$ & $0.000152$ & $2.71\times$ \\
1024 &$0.000945$ & $0.000838$ & $1.13\times$ &
$0.000914$ & $0.000519$ & $1.76\times$ &
$0.000702$ & $0.000645$ & $1.09\times$ &
$0.000046$ & $0.000054$ & $0.852\times$ \\
  \end{tabular}
  \caption{MRSE the four \emph{Sportscar} scenes at various power-of-4
    numbers of pixel samples. For most scenes and most sampling rates, MRSE is lower
    with the SZ sampler than with the Sobol sampler.
    For \emph{Sky} SZ sampling yields much lower error at 16 and 256spp but slightly
    higher at 64 and 1024spp.
  }
  \label{tab:rendered-mrse}
\end{table*}

\begin{table*}[tb]
  \centering
  \begin{tabular}{r|rrr|rrr|rrr|rrr}
 & \multicolumn{3}{c|}{Hangar Interior} & \multicolumn{3}{c|}{Cayley Interior} & \multicolumn{3}{c|}{Empty Warehouse} & \multicolumn{3}{c}{Sky} \\
 spp & \multicolumn{1}{c}{Sobol} & \multicolumn{1}{c}{SZ} & \multicolumn{1}{c|}{Diff.} &
 \multicolumn{1}{c}{Sobol} & \multicolumn{1}{c}{SZ} & \multicolumn{1}{c|}{Diff.} &
 \multicolumn{1}{c}{Sobol} & \multicolumn{1}{c}{SZ} & \multicolumn{1}{c|}{Diff.} &
 \multicolumn{1}{c}{Sobol} & \multicolumn{1}{c}{SZ} & \multicolumn{1}{c}{Diff.} \\
\hline
16 &$0.101$ & $0.0986$ & $0.00245$ &
$0.0852$ & $0.0795$ & $0.00573$ &
$0.0762$ & $0.0712$ & $0.00497$ &
$0.0413$ & $0.0297$ & $0.0117$ \\
64 &$0.0497$ & $0.0473$ & $0.00237$ &
$0.0379$ & $0.0348$ & $0.00314$ &
$0.0351$ & $0.0323$ & $0.00285$ &
$0.0126$ & $0.0122$ & $0.000409$ \\
256 &$0.024$ & $0.0255$ & $-0.00154$ &
$0.0192$ & $0.0194$ & $-0.000228$ &
$0.017$ & $0.0167$ & $0.000275$ &
$0.00606$ & $0.00622$ & $-0.000159$ \\
1024 &$0.0137$ & $0.0151$ & $-0.00149$ &
$0.0111$ & $0.0119$ & $-0.000744$ &
$0.00934$ & $0.0107$ & $-0.00132$ &
$0.0029$ & $0.00482$ & $-0.00193$ \\
  \end{tabular}
  \caption{\FLIP error for the test scenes at various power-of-4 numbers of pixel samples.
    \FLIP error is consistently lower with the SZ sampler at 16 and 64 samples per pixel; at 256
    and 1024 samples per pixel, the image is nearly converged and \FLIP reports nearly equal error for both samplers. 
  }
  \label{tab:rendered-flip}
\end{table*}

\end{document}